%% file: paper.tex
\documentclass[journal,twoside,twocolumn,letterpaper]{IEEEtran}

\usepackage{ifpdf}
\usepackage{color}
\usepackage[latin1]{inputenc}
\usepackage{cite}
\usepackage{mathtools}
\usepackage{amssymb}
\usepackage{url,cite}

\usepackage{subfigure}

\usepackage{tikz}
\usetikzlibrary{decorations.pathmorphing} 
\usetikzlibrary{fit}					
\usetikzlibrary{backgrounds}	
\usetikzlibrary{shapes}
\usetikzlibrary{positioning}
\usetikzlibrary{shadows}
\usetikzlibrary{decorations.pathmorphing}
\usetikzlibrary{decorations.shapes}
\usepackage{multirow}
\usepackage{dcolumn}
\usepackage{pgfplots}
\usepackage{xcolor}
\pgfplotsset{compat=newest}

\newtheorem{theorem}{Theorem}
\newtheorem{lemma}{Lemma}
\newtheorem{definition}{Definition}
\newtheorem{corollary}{Corollary}

\begin{document}
\allowdisplaybreaks

\pagestyle{headings}
\title{Auto-Correlation and Coherence Time of Interference in Poisson Networks}
\author{Udo Schilcher\thanks{Udo Schilcher is with Lakeside Labs GmbH, Klagenfurt, Austria, email: \texttt{schilcher@lakeside-labs.com}.}, 
			Jorge F. Schmidt, Mahin K. Atiq, and Christian Bettstetter
			\thanks{Jorge F. Schmidt, Mahin K. Atiq, and Christian Bettstetter are with the University of Klagenfurt, Networked and Embedded Systems, Klagenfurt, Austria.}
}

\markboth{}{}
\thispagestyle{empty}
\makeatletter
\newcommand{\nvast}{\bBigg@{3}}
\makeatother

\newcommand{\etal}{{et al.\;}}

\newcommand{\prob}[1]{{\mathbb{P}\!\left[{#1}\right]}}
\newcommand{\expect}[1]{{{\mathbb{E}}\!\left[{#1}\right]}}
\newcommand{\var}[1]{{{\rm var}\!\left[{#1}\right]}}
\newcommand{\cov}[2]{{{\rm cov}\!\left[{#1},{#2}\right]}}
\newcommand{\cor}[2]{{\rho\!\left[{#1},{#2}\right]}}

\newcommand{\eulergamma}{\gamma_{\mathrm{eul}}}
\newcommand{\Ei}[1]{E_i\left(#1\right)}

\newcommand{\R}{\mathbb{R}}
\newcommand{\N}{\mathbb{N}}
\newcommand{\dd}{{\rm d}}
\newcommand{\uniform}{\mathbb{U}(0,1)}

\newcommand{\ppp}{\Phi}
\newcommand{\interf}{I}
\newcommand{\fad}[2]{h_{#2}^2}
\newcommand{\fadsq}[2]{h_{#2}^4}
\newcommand{\ploss}[1]{\ell\left(\left\|#1\right\|\right)}
\newcommand{\plossn}[1]{\ell_{#1}}
\newcommand{\plosssc}[1]{\ell\left(#1\right)}
\newcommand{\plosssq}[1]{\ell^2(\|#1\|)}
\newcommand{\plossxy}{\ell_{xy}}
\newcommand{\plossx}{\ell_{x}}
\newcommand{\plossy}{\ell_{y}}

\newcommand{\channel}[2]{\fad{#1}{#2}\,\plossn{#1}}
\newcommand{\channelx}{\fad{t}{x}\,\plossx}

\newcommand{\tx}[2]{\gamma_{#2}}
\newcommand{\txone}[1]{\gamma^{\rm I}_{#1}}
\newcommand{\txtwo}[1]{\gamma^{\rm II}_{#1}}
\newcommand{\prdensii}[1]{\rho^{(2)}(#1)}
\newcommand{\dens}{\lambda}

\newcommand{\txpwr}{\kappa}
\newcommand{\rxpwr}{p_\mathrm{RX}}
\newcommand{\nakm}{m}
\newcommand{\plc}{\alpha}
\newcommand{\txp}{p}
\newcommand{\txpc}{{q}}
\newcommand{\txpd}{\mu}

\newcommand{\chlen}{c}
\newcommand{\msglen}{d}

\newcommand{\ti}{t}
\newcommand{\delay}{\tau}
\newcommand{\snd}[1]{S(#1)}

\newcommand{\pa}{p_1}
\newcommand{\pab}{p_{12}}
\newcommand{\papa}{p_{1/1}}
\newcommand{\papb}{p_{1/2}}

\newcommand{\areac}{A_c}
\newcommand{\areax}{A_x}
\newcommand{\areay}{A_y}

\newcommand{\cir}{\mathcal{C}}

\newcommand{\expectppp}[1]{{{\mathbb{E}}\!\left[{#1}\right]}}
\newcommand{\expecth}[1]{{{\mathbb{E}}\!\left[{#1}\right]}}
\newcommand{\expecttx}[1]{{{\mathbb{E}}\!\left[{#1}\right]}}

\newcommand{\gap}{g}

\newcommand{\speed}{\bar{v}}
\newcommand{\mobstep}{\omega}
\newcommand{\mob}{{\omega_\delay}}
\newcommand{\expectmob}[1]{{{\mathbb{E}}\!\left[{#1}\right]}}

\newcommand{\cohti}{\delay_c}
\newcommand{\cohth}{\theta}

\sloppy

\maketitle
\thispagestyle{empty}

\input{abstract.tex}

\input{intro_new.tex}

\input{model.tex}

\input{cor.tex}

\input{src.tex}

\input{coherence.tex}

\input{conclusion.tex}

\section*{Acknowledgments}
This work has been supported by the Austrian Science Fund (FWF) under grant P24480-N15 (Dynamics of Interference in Wireless Networks) and by the K-project DeSSnet.
The K-Project DeSSnet is funded within the context of COMET -- Competence Centers for Excellent Technologies by the Austrian Ministry for Transport, Innovation and Technology (BMVIT), the Federal Ministry for Digital and Economic Affairs (BMDW), and the federal states of Styria and Carinthia. The program is conducted by the Austrian Research Promotion Agency (FFG).


\input{bios.tex}

\end{document}

%% file: abstract.tex
\begin{abstract}
The dynamics of interference over space and time influences the performance of wireless communication systems, yet its features are still not fully understood. 
This article analyzes the temporal dynamics of the interference in Poisson networks accounting for three key correlation sources: the location of nodes, the wireless channel, and the network traffic. 
We derive expressions for the auto-correlation function of interference. These are presented as a framework that enables us to arbitrarily combine the three correlation sources to match a wide range of interference scenarios. 
We then introduce the interference coherence time --- analogously to the well-known channel coherence time --- and analyze its features for each correlation source. 
We find that the coherence time behaves very different for the different interference scenarios considered and depends on the network parameters. 
Having accurate knowledge of the coherence time can thus be an important design input for protocols, e.g., retransmission and medium access control.
\end{abstract}
\begin{IEEEkeywords}
Interference, auto-correlation, coherence time, Poisson point process.
\end{IEEEkeywords}

%% file: intro_new.tex
\section{Introduction}\label{sec:intro}
\subsection{Motivation}

The performance analysis of mobile communication and computing networks must model the interference caused by nodes on other nodes. Common models describe the average value or distribution of interference at a given receiver. More accurate modeling also considers the {\it interference dynamics\/}, i.e., the way interference changes over time and space~\cite{andrews17:mmWave, 8115171, haenggi13:book, schilcher15:tit, haenggi13:div-poly}. Questions of interest are: How rapidly does interference change? What is the correlation of interference? Which parameters influence this correlation? Answers to these questions are particularly useful for the design of retransmission protocols and diversity schemes. Various researchers addressed these questions, taking different views.

A first view adopts measures like outage probability (or its counterpart, the coverage probability) for the purpose of protocol performance evaluation. This requires knowledge of the probability distribution of the signal-to-interference ratio (SIR), which in many cases is easier to calculate than some interference statistics~\cite{haenggi13:book}. It is, for example, possible to analyze the performance degradation or improvement of cooperative relaying~\cite{tanbourgi14:nakagami,tanbourgi14:mrc,crismani14:tvt,schilcher13:mswim} or MIMO~\cite{7880697,haenggi13:div-poly} impacted by interference.

A second view is to use interference dynamics to determine interference statistics, i.e., to establish the probability distribution of interference power in two or more points in time and/or space. Such multidimensional probability distribution is best represented by its probability density function. Unfortunately, for interference this function does not exist in closed form unless we consider scenarios with restricted network parameter values. Hence, we have to resort to characteristic functions, which are less flexible to handle but still valuable as they allow to calculate the moments of the distribution~\cite{schilcher15:tit,haenggi13:book} and other values.

Finally, a third view is to quantify the dependence of interference powers at different points in time and/or space. In other words, it addresses the question as to how much information is gained on the interference power (at a certain point in time and space) if we know interference power values close-by or from the past. Mathematically, this dependence is expressed in terms of correlation, with one application being interference prediction~\cite{atiq17:mswim}: The auto-correlation function of the interference determines the order of the time series associated to the interference evolution over time, and can be the basis for designing a predictor.

The auto-correlation function of interference is still unknown. The reason is that most work on interference dynamics only considers the node locations as source of interference correlation. This assumption has two important consequences: First, the auto-correlation of interference at a given point in space is independent of the time lag $\tau$, i.e., it is a constant function~\cite{ganti09:interf-correl} not especially interesting to adopt in network performance analysis. Second, the resulting interference correlation may be inaccurate as important other sources of correlation are neglected. Two such additional sources of correlation are the data traffic and the wireless channel~\cite{schilcher12:tmc}. When considering these sources, the auto-correlation becomes non-constant and thus turns into an important tool for wireless network modeling and analysis.

In the article at hand, we generalize the known expressions for temporal correlation of interference in consecutive time slots (our previous work~\cite{schilcher12:tmc}) to arbitrary time lags $\tau$. In other words, we calculate the \textit{auto-correlation function} (ACF) of interference for this $\tau$. Our work is based on the commonly used Poisson model, which means that the nodes of the network are distributed in space according to a Poisson point process. Random access to the wireless channel without channel sensing is used. As part of this work, we present modular expressions that allow different fading and traffic models to be incorporated by simply substituting a corresponding expression. In particular, we derive ACF expressions both individually for the three sources of correlation and for particular combinations of these sources. Furthermore, we analyze the \textit{coherence time} of interference, which is the time until the interference correlation falls below a given threshold. We are able to derive closed form expressions in some cases and rely on numerical analysis in others.

\subsection{Related work}
There are several results on the temporal correlation of interference in Poisson networks (e.g.,~\cite{ganti09:interf-correl,6697936,haenggi13:div-poly,6331038,schilcher12:tmc,schilcher13:scc}). All these results, however, consider only the node locations as source of correlation and assume a static network without mobility. Because of these assumptions the temporal correlation does not change over time, i.e., it does not depend on the time lag $\delay$ between the two instants under consideration. As a natural consequence, neither auto-correlation nor coherence time is considered.

Furthermore, results are available on the stochastic dependence of interference, which is typically expressed as the joint outage probability of several transmissions (e.g.,~\cite{schilcher15:tit,tanbourgi14:mrc,tanbourgi14:nakagami,crismani14:tvt,net:Haenggi09jsac,haenggi13:book}). In all these publications, again the joint outage probability is independent of the time instant of the transmissions, which is a consequence of assuming that the node locations are the only source of correlation and that nodes are not mobile.

Important exceptions to these limitations are the results by Gong and Haenggi~\cite{gong14:tmc, gong11:icc}. They also consider the node locations as the sole source of interference correlation but use a mobile network. It is the mobility that makes the temporal correlation to decrease for longer time lags $\delay$. 
The analysis is performed for four different stochastic mobility models including Brownian motion (also adopted in the article at hand). The dependence of the correlation on the system parameters is analyzed with special focus on the average speed of the nodes. The evolution of the correlation in terms of the time lag is not extensively analyzed in those two publications and thus no notion of coherence time is considered. 

The three sources of interference investigated in the article at hand are used in~\cite{schilcher12:tmc}, where all 27 possible combinations of them are systematically addressed. However, the temporal correlation is only calculated for consecutive time slots, which provides no insights on the time it takes to reach low or negative correlation values.

\subsection{Main contributions}
We investigate the temporal dynamics of the interference in terms of its auto-correlation function and coherence time. We do so by accounting for three key sources of interference correlation: the node locations, the wireless channel, and the network traffic.
Prior to this work, temporal correlation was only known for consecutive time slots~\cite{schilcher12:tmc}, and for longer lags, only when considering the node locations as sole source of correlation. Only two possible outcomes apply: If there is no mobility among the nodes, the temporal correlation does not change over time~\cite{ganti09:interf-correl}, while for mobility it decreases~\cite{gong14:tmc} with increasing lags.

Our contributions can be summarized as follows:
\begin{itemize}
	\item We are the first to derive expressions for the auto-correlation function of interference in Poisson networks for the three key sources of correlation. These provide insights on the temporal dynamics of interference far beyond existing results, and are relevant for facilitating the exploitation of the temporal features of interference in emerging wireless systems. 
	\item The mathematical framework provided enables us to combine the three sources of correlation into a general expression for the temporal correlation of interference. Hence, unlike in previous work~\cite{schilcher12:tmc}, it is no longer needed to address all possible combinations of the sources individually. Results are much more flexible and easier to apply in further work.
	\item The analysis of the interference coherence time is further contributing to the practical applicability of our theoretical work. It provides an answer to the question as to how long a retransmission protocol has to wait for an uncorrelated channel.
\end{itemize}
Our expressions also cover the fundamental and well-known channel coherence time without interference, which corresponds to taking only the channel into account as a source of correlation. Under this assumption, the interference coherence time and the channel coherence time are equal. 

The rest of the article is organized as follows: Section~\ref{sec:model} introduces the modeling assumptions. Section~\ref{sec:cor} provides general expressions for the temporal correlation of interference, which need specific sub-expression that model the different sources of interference correlation to be substituted. In Section~\ref{sec:src} the expressions that have to be substituted into the equations for temporal interference correlation are derived. Based on these results, the coherence time of interference is defined, and expressions for it are derived and used to analyze its features for different parameters in Section~\ref{sec:coh}. Finally, Section~\ref{sec:con} concludes the paper.

%% file: model.tex
\section{System model}\label{sec:model}
\subsection{Node location and traffic}
We consider a Poisson network: a wireless network consisting of nodes randomly located in a plane according to a Poisson point process (PPP) $\ppp$ on $\R^2$.
The medium access opportunities are arranged into time slots. In each slot, every idle node decides independently from other nodes whether or not to start a new transmission. The duration of this transmission is $\msglen\in\N$ time slots (message length), which is constant for all nodes and over time.
We intend to have on average a fraction $\txp$ of all nodes in $\ppp$ start a new transmission in each slot.
Since only idle nodes start new transmissions, they adopt a sending probability of $\frac{\txp}{1-\txp(\msglen-1)}$.
Let $\txpc$ denote the probability that a node stays idle, i.e., $\txpc=1-\frac{\txp}{1-\txp(\msglen-1)}$ and $S_t$ denote the set of all sending nodes at time $\ti$.
The expected fraction of nodes sending in a given time slot $t$ is thus $\txpd=|S_t|=\txp\,\msglen$, which we refer to as the traffic intensity.

\subsection{Node mobility}
Both static and mobile nodes are investigated.
For mobile nodes, we let $\speed$ denote the average speed of all nodes and consider two mobility models: linear mobility and time-discrete Brownian motion.
The linear mobility model assumes the location of each node $x$ at times $\ti$ and $\ti+\delay$ to change in a random direction with $|x_{\ti}-x_{\ti+\delay}|=\speed\delay$, i.e., the distance increases linearly with time and we set $\mob=\delay$. This can be considered to be a reasonable model for time spans covering the duration of a few time slots, as the direction of movement and speed does not change significantly within the timescale of a slot in a practical system.

The location of a node $x$ with Brownian motion at time $\ti+1$ is~\cite{gong14:tmc,gong11:icc}
\begin{equation}
x_{\ti+1}=x_{\ti}+\speed\mobstep_{\ti}\:,
\end{equation}
where $\mobstep_\ti$ is a two-dimensional Gaussian random variable $\mobstep_\ti\sim N(\vec{0},\Sigma)$ with covariance matrix 
\begin{equation}
\Sigma=
\begin{pmatrix}
	0 & \sqrt{\frac{2}{\pi}}\\
	\sqrt{\frac{2}{\pi}} & 0
\end{pmatrix}\:.
\end{equation}
The random variables $\mobstep$ are i.i.d. for each time $\ti$. Hence, the location after $\delay$ slots is
\begin{equation}
\mob=\sum_{\ti=1}^\delay x_t\stackrel{d}{=}\sqrt{\delay}\,\mobstep_0\:,
\end{equation}
where $\stackrel{d}{=}$ denotes the equality in distribution.

\textbf{Remark:} The homogeneity of the PPP describing the node locations is not altered by these two mobility models. At any time $\ti$ the location of nodes forms a PPP with intensity $\dens$. Note that this preservation of homogeneity is not provided by many other mobility models, for example, for the random waypoint model (nodes asymptotically concentrate in the center of the deployment area~\cite{bettstetter03:tmc}).

\subsection{Wireless channel}
The wireless channel is modeled by a distance dependent path loss and multi-path fading accounting for reflections, diffraction, and other small-scale propagation effects.
The signal power at a receiver $x$ from an active sender $y$, with $x,y\in\R^2$ is
\begin{equation}
\rxpwr=\txpwr\,h_\ti^2\,\plossxy\:.
\end{equation}
In this equation, $\txpwr$ is the sending power of $y$, which we consider to be the same for all nodes in the network.
$h_\ti$ models Nakagami-$\nakm$ block fading at time $\ti$ and node $x$, i.e., $h_\ti^2$ is gamma distributed according to $h_\ti^2\sim\Gamma(\nakm,\nakm)$. This implies $\expect{h_\ti^2}=1$ and $\expect{h_\ti^4}=\frac{\nakm+1}{\nakm}
$. The temporal aspect of fading is modeled in the following way~\cite{schilcher12:tmc}: 
We consider a block fading model, where the channel is assumed to remain constant over a duration of $\chlen\in\N$ time slots after which it changes to an independent state, i.e., a random experiment is carried out independently of the previous channel state to establish the new channel state. This model for the temporal behavior is of widespread use. It matches well practical systems where the signal timing is usually designed to approximately meet this condition for easing the channel state acquisition and equalization tasks.
Finally, $\plossxy=\ploss{y-x}$ denotes a non-singular distance dependent path loss, for which we adopt 
\begin{equation}
\ploss{y-x}=\min(1, \|y-x\|^{-\plc})
\end{equation}
with a path loss exponent $\plc>2$. Let $\plossx=\ploss{x-o}=\ploss{x}$ be the path loss from a node $x$ to the origin $o$.

\subsection{Interference}
Interference at time $\ti$ is measured at the origin $o$ of the plane $\R^2$, which is equal to the interference experienced by a typical node of the network due to Slivnyak's theorem. Its power is the sum of the signal powers of all sending nodes in the network (besides the intended signal from an specific sender, which is not considered in this work) yielding
\begin{equation}
\interf_\ti=\sum_{x\in\ppp}\txpwr\,h_\ti^2\,\plossx\,\gamma_\ti\:,
\end{equation}
where $\gamma_\ti$ is a Bernoulli random variable indicating whether node $x$ is sending ($\gamma_\ti=1$) at time $\ti$ or not ($\gamma_\ti=0$).

\subsection{Classification of correlation sources}\label{ssec:cases}
We consider three sources of correlation of interference: node locations, wireless channel (i.e. correlated fading), and traffic. For each of them, there are three possible options, denoted by a triplet $(i,j,k)\in\{0,1,2\}^3$:
\begin{itemize}
	\item They are constant or the correlation is not considered (denoted by $0$).
	\item They are random but uncorrelated (denoted by $1$).
	\item They are random and correlated (denoted by $2$).
\end{itemize}
This leads to $27$ different cases that have been introduced in more detail and analyzed with respect to temporal correlation of interference in~\cite{schilcher12:tmc}.

%% file: cor.tex
\section{Derivation of the auto-correlation of interference}\label{sec:cor}

\begin{table*}[t]
\begin{center}
\caption{Summary of results without mobility: Auto-correlation and coherence time of interference. For the coherence time expressions, all resulting non-integer values have to be rounded to the next higher integer. The symbol '-' denotes that we have no expression for this case.\label{tab:results}}
\begin{tabular}{ c c c | c | c }
 Locations & Channel & Traffic & Auto-correlation function & Coherence time\\
 $i$ & $j$ & $k$ & $\rho(\tau)$ & $\cohti$ \\
\hline
 $0$  &  $0$  &  $0$  & undefined & undefined\\
 $0$  &  $0$  &  $1$  & $0$ & $1$\\
 $0$  &  $0$  &  $2$  & $\frac{\txpc^\tau+\mu-1}{\mu}$ for $\tau<d$ & $\frac{\log(1-\mu)}{\log\txpc}$\\
 $0$  &  $1$  &  $0$  & $0$ & $1$\\
 $0$  &  $1$  &  $1$  & $0$ & $1$\\
 $0$  &  $1$  &  $2$  & $\frac{m(\mu-1)(\txpc^\tau+\mu-1)}{\mu(m(\mu-1)-1)}$ for $\tau<d$ & $\frac{\log(1-\mu)}{\log\txpc}$\\
 $0$  &  $2$  &  $0$  & $1-\frac{\tau}{c}$ for $\tau\leq c$ & $c$\\
 $0$  &  $2$  &  $1$  & $\frac{p(c-\tau)}{c(1+m-mp)}$ for $\tau<c$ & $c$\\
 $0$  &  $2$  &  $2$  & $\frac{\left(\txpc^\tau(\mu-1)-2\mu+1\right)(\tau-c(m+1))-cm\mu^2}{c\mu(1+m-m\mu)}$ for $\tau<\min(c,d)$ & -\\
 $1$  & $0,1$, or $2$ & $0,1$, or $2$ & $0$ & $1$\\
 $2$  &  $0$  &  $0$  &   $1$ & $\infty$ \\
 $2$  &  $0$  &  $1$  &   $p$ & $\infty$ \\
 $2$  &  $1$  &  $0$  &   $1/2$ & $\infty$ \\
 $2$  &  $1$  &  $1$  &   $p/2$ & $\infty$ \\
 $2$  &  $0,1$ & $2$  & $\frac{\txpc^\tau (1-\mu)+2\mu-1}{\mathbb{E}[h^4]\mu}$ for $\tau<d$ & -\\
 $2$  &  $2$  &  $0$  & $1-\frac{\tau}{c(m+1)}$ for $\tau<c$ & -\\
 $2$  &  $2$  &  $1$  & $p\left(1-\frac{\tau}{c(m+1)}\right)$ for $\tau<c$ & -\\
 $2$  &  $2$  &  $2$  & $\frac{\left(\txpc^\tau(\mu-1)-2\mu+1\right)(\tau-c(m+1))}{c\mu(m+1)}$ for $\tau<\min(c,d)$ & -
\end{tabular}
\end{center}
\end{table*}

We derive in the following paragraphs general expressions for the correlation of interference. These expressions are further specialized for selected interference scenarios in the next section.
Interference correlation is measured in terms of Pearson's correlation coefficient
\begin{eqnarray}
\cor{\interf_{\ti_1}}{\interf_{\ti_2}}&=&\frac{\cov{\interf_{\ti_1}}{\interf_{\ti_2}}}{\sqrt{\var{\interf_{\ti_1}}\var{\interf_{\ti_2}}}}\\\nonumber
&=&\frac{\cov{\interf_{\ti_1}}{\interf_{\ti_2}}}{\var{\interf}}\:,
\end{eqnarray}
where $\cov{\interf_{\ti_1}}{\interf_{\ti_2}}=\expect{\interf_{\ti_1}\interf_{\ti_2}}-\expect{\interf_{\ti_1}}\,\expect{\interf_{\ti_1}}$ denotes the covariance of interference at different time instants $\ti_1$ and $\ti_2$, and $\var{\interf}=\var{\interf_{\ti_1}}=\var{\interf_{\ti_2}}$ denotes the variance of interference, which is constant over time due to the stationarity of the processes involved. The time lag from $\ti_1$ to $\ti_2$ is denoted by $\delay=\ti_2-\ti_1$.

In the derivation of the general expressions, we distinguish between random and known node locations. In the former case we consider node locations as sources of interference correlation (cases $(2,j,k)$) while in the latter we do not consider them as a source (cases $(0,j,k)$).

A summary of the auto-correlation and the coherence time of interference without mobility is presented in Table~\ref{tab:results}. The expressions in this table only hold for small time lags $\tau<c$ and $\tau<d$. The more general expressions for arbitrary $\tau$, and for mobility, are too long for the table but are available in the following sections. Coherence time results are computed by solving $\rho(\tau)=0$ for $\tau$. For the cases with '-' this can also be solved, but leads to expressions that have $\tau>\chlen$ or $\tau>\msglen$ for all parameters, which violates the presumptions for the simplified expressions of the auto-correlation functions presented in this table. Adopting the full expression for the auto-correlation does not lead to closed form expressions for coherence time.
Note that when substituting $\tau=1$ and $m=1$ to the expressions in the table, the correlation results correspond to that in Table~1 of~\cite{schilcher12:tmc} with the following exception:
In the cases $(i,2,1)$ and $(i,2,2)$ there is a difference due to a difference in the modeling assumptions. Here we assume that the channel of any pair of nodes is changing every $c$ slots while in~\cite{schilcher12:tmc} the channel changes $c$ slots after a transmission occurred.

\subsection{Random node locations}
\begin{theorem}[Correlation for cases $(2,j,k)$]\label{th:case2xx}
The temporal correlation of interference between time instants $\ti_1$ and $\ti_2$ considering the node locations as a source of correlation is
\begin{eqnarray}\label{eq:cor2xxmob}
\lefteqn{\cor{\interf_{\ti_1}}{\interf_{\ti_2}}=}\\\nonumber
&=&\frac{\expect{\fad{x}{\ti_1}\fad{x}{\ti_2}}\,\expect{\tx{x}{\ti_1}\tx{x}{\ti_2}}}{\txpd\,\expect{\fadsq{x}{\ti}}}\cdot\frac{\int_{\R^2}\plossx\,\expectmob{\ploss{x+\speed\mob}}\,\dd x}{\int_{\R^2}\plossx^2\,\dd x}\:.
\end{eqnarray}
\end{theorem}
\begin{IEEEproof}
The expected value of interference is 
\begin{eqnarray}
\expect{\interf}&=&\mathbb{E}_{\ppp,h,\gamma}{\sum_{x\in\ppp}\channelx\tx{x}{\ti}}\\\nonumber
&=&\mathbb{E}_\ppp\sum_{x\in\ppp}\mathbb{E}_h[\fad{x}{\ti}]\,\plossx\,\,\mathbb{E}_\gamma[\tx{x}{\ti}]\\\nonumber
&\stackrel{(a)}{=}&\msglen\txp\,\dens\int_{\R^2}\plossx\,\dd x\\\nonumber
&=&\txpd\,\dens\,\frac{\plc\pi}{\plc-2}\:,
\end{eqnarray}
where $(a)$ holds due to Campbell's theorem~\cite{haenggi13:book}, $\expect{\fad{x}{\ti}}=1$ for all $x\in\ppp$ and $\ti\in\N$ and $\expect{\tx{x}{\ti}}=\msglen\txp$.
The indices for the expectation operator $\mathbb{E}$ indicate the random variables involved and will be omitted in the future for shorter expressions.
Aiming for the covariance, we calculate
\begin{eqnarray}\label{eq:pr:2xxcoexp}
\lefteqn{\expect{\interf_{\ti_1}\interf_{\ti_2}}=}\\\nonumber
&=&\expect{\sum_{x\in\ppp}\channel{x}{\ti_1}\tx{x}{\ti_1}\sum_{y\in\ppp}\tilde{h}_{\ti_2}^2\ploss{y+\speed\mob}\tilde{\gamma}_{\ti_2}}\\\nonumber
&=&\expect{\sum_{x\in\ppp}\fad{x}{\ti_1}\fad{x}{\ti_2}\plossx\,\ploss{x+\speed\mob}\tx{x}{\ti_1}\tx{x}{\ti_2}}\\\nonumber
&&+\,\expect{\sum_{\stackrel{x,y\in\ppp}{x\neq y}}\fad{x}{\ti_1}\tilde{h}_{\ti_2}^2\plossx\,\ploss{y+\speed\mob}\tx{x}{\ti_1}\tilde{\gamma}_{\ti_2}}\:,
\end{eqnarray}
where we introduce $\tilde{h}_{\ti}$ and $\tilde{\gamma}_{\ti}$ to denote the fading coefficient and sending indicator of node $y$ at time $\ti$, respectively. The first of the two expected values of~\eqref{eq:pr:2xxcoexp} yields
\begin{eqnarray}
\lefteqn{\expect{\sum_{x\in\ppp}\fad{x}{\ti_1}\fad{x}{\ti_2}\plossx\,\ploss{x+\speed\mob}\tx{x}{\ti_1}\tx{x}{\ti_2}}=}\\\nonumber
&=&\expectppp{\sum_{x\in\ppp}\expect{\fad{x}{\ti_1}\fad{x}{\ti_2}}\plossx\,\expectmob{\ploss{x+\speed\mob}}\expect{\tx{x}{\ti_1}\tx{x}{\ti_2}}}\\\nonumber
&=&\expect{\fad{x}{\ti_1}\fad{x}{\ti_2}}\,\expect{\tx{x}{\ti_1}\tx{x}{\ti_2}}\,\dens\int_{\R^2}\plossx\,\expectmob{\ploss{x+\speed\mob}}\dd x\:,
\end{eqnarray}
and the second gives
\begin{eqnarray}
\lefteqn{\expect{\sum_{\stackrel{x,y\in\ppp}{x\neq y}}\fad{x}{\ti_1}\tilde{h}_{\ti_2}^2\plossx\,\expectmob{\ploss{y+\speed\mob}}\tx{x}{\ti_1}\tilde{\gamma}_{\ti_2}}=}\\\nonumber
&\stackrel{(a)}{=}&\dens^2\int_{\R^2}\int_{\R^2}\expect{\fad{x}{\ti_1}\tilde{h}_{\ti_2}^2}\plossx\ploss{y+\speed\mob}\expect{\tx{x}{\ti_1}\tilde{\gamma}_{\ti_2}}\,\dd x\,\dd y\\\nonumber
&\stackrel{(b)}{=}&\left(\txpd\,\dens\int_{\R^2}\plossx\,\dd x\right)^2\\\nonumber
&=&\expect{\interf}^2\:.
\end{eqnarray}
In $(a)$ we split the expected value for the different independent random variables and apply Campbell's theorem. In $(b)$ we use $\expect{\fad{x}{\ti_1}\tilde{h}_{\ti_2}^2}=1$, $\expect{\tx{x}{\ti_1}\tilde{\gamma}_{\ti_2}}=\txpd^2$ and the stationarity of the PPP.
Hence, the covariance is
\begin{eqnarray}\label{eq:pr:2xxcov}
\lefteqn{\cov{\interf_{\ti_1}}{\interf_{\ti_2}}=}\\\nonumber
&=&\expect{\interf_{\ti_1}\interf_{\ti_2}}-\expect{\interf_{\ti_1}}\expect{\interf_{\ti_2}}\\\nonumber
&=&\expect{\fad{x}{\ti_1}\fad{x}{\ti_2}}\,\expect{\tx{x}{\ti_1}\tx{x}{\ti_2}}
\,\dens\,\int_{\R^2}\plossx\,\expectmob{\ploss{x+\speed\mob}}\,\dd x\:.
\end{eqnarray}
The values of $\expect{\fad{x}{\ti_1}\fad{x}{\ti_2}}$ and $\expect{\tx{x}{\ti_1}\tx{x}{\ti_2}}$ are characterizing the contribution of the wireless channel and the traffic to the correlation of interference, respectively. They depend on the values of $j,k$ of the case $(2,j,k)$ under consideration.

The variance is obtained by setting $\ti_1=\ti_2$ in the above derivations yielding
\begin{equation}\label{eq:pr:2xxvar}
\var{\interf}=\cov{\interf}{\interf}=\txpd\,\expect{\fadsq{x}{\ti}}\,\dens\,\int_{\R^2}\plossx^2\,\,\dd x\:.
\end{equation}
Dividing~\eqref{eq:pr:2xxcov} by~\eqref{eq:pr:2xxvar} yields the result.
\end{IEEEproof}

\begin{corollary}[Correlation for cases $(2,j,k)$ without mobility]\label{cor:case2xx}
The temporal correlation of interference between time instants $\ti_1$ and $\ti_2$ considering the node locations as sources of interference correlation and having no mobility ($\speed=0$) is
\begin{equation}\label{eq:cor2xx}
\cor{\interf_{\ti_1}}{\interf_{\ti_2}}=\frac{\expect{\fad{x}{\ti_1}\fad{x}{\ti_2}}\,\expect{\tx{x}{\ti_1}\tx{x}{\ti_2}}}{\txpd\,\expect{\fadsq{x}{\ti}}}\:.
\end{equation}
\end{corollary}
\begin{IEEEproof}
Substituting $\speed=0$ into~\eqref{eq:cor2xxmob} yields the result.
\end{IEEEproof}

\subsection{Known node locations}
\begin{theorem}[Correlation for cases $(0,j,k)$]\label{th:case0xx}
The temporal correlation of interference between time instants $\ti_1$ and $\ti_2$ if neglecting the node locations as sources of interference correlation is
\begin{equation}
\cor{\interf_{\ti_1}}{\interf_{\ti_2}}=\frac{\expect{\fad{x}{\ti_1}\fad{x}{\ti_2}}\,\expect{\tx{x}{\ti_1}\tx{x}{\ti_2}}-\txpd^2}{\txpd\,\expect{\fadsq{x}{\ti}}-\txpd^2}\:.
\end{equation}
\end{theorem}
\begin{IEEEproof}
The covariance of interference is
\begin{eqnarray}
\cov{\interf_{\ti_1}}{\interf_{\ti_2}}
\hspace{-1mm}&=&\hspace{-1mm}\expectppp{\sum_{x\in\ppp}\sum_{y\in\ppp}\cov{\channel{x}{\ti_1}\tx{x}{\ti_1}}{\tilde{h}_{\ti_2}^2\,\plossy\,\tilde{\gamma}_{\ti_2}}}\hspace{4mm}\\\nonumber
&=&\hspace{-1mm}\expectppp{\sum_{x\in\ppp}\cov{\channel{x}{\ti_1}\tx{x}{\ti_1}}{\channel{x}{\ti_2}\tx{x}{\ti_2}}}\\\nonumber
&&\hspace{-1mm}+\,\expectppp{\sum_{\stackrel{x,y\in\ppp}{x\neq y}}\cov{\channel{x}{\ti_1}\tx{x}{\ti_1}}{\tilde{h}_{\ti_2}^2\,\plossy\,\tilde{\gamma}_{\ti_2}}}\:.
\end{eqnarray}
The covariance in the second sum is always zero as the arguments are stochastically independent.
The covariance in the first sum yields
\begin{eqnarray}\label{eq:pr:0xxcov}
\lefteqn{\expectppp{\sum_{x\in\ppp}\cov{\channel{x}{\ti_1}\tx{x}{\ti_1}}{\channel{x}{\ti_2}\tx{x}{\ti_2}}}=}\\\nonumber
&\stackrel{(a)}{=}&\mathbb{E}\Bigg[\sum_{x\in\ppp}\expect{\fad{x}{\ti_1}\fad{x}{\ti_2}}\,\plossx^2\,\,\expect{\tx{x}{\ti_1}\tx{x}{\ti_2}}
-\big(\expect{\fad{x}{\ti}}\,\plossx\,\,\expect{\tx{x}{\ti}}\big)^2\Bigg]\\\nonumber
&\stackrel{(b)}{=}&\left(\expect{\fad{x}{\ti_1}\fad{x}{\ti_2}}\,\expect{\tx{x}{\ti_1}\tx{x}{\ti_2}}-\txpd^2\right)
\,\dens\,\int_{\R^2}\plossx^2\,\,\dd x\:,
\end{eqnarray}
where in $(a)$ we calculated the covariance with $\cov{X}{Y}=\expect{XY}-\expect{X}\expect{Y}$ and in $(b)$ we substituted $\expect{\fad{x}{\ti}}=1$ and $\expect{\tx{x}{\ti}}=\txpd$.
Similar to the proof of Theorem~\ref{th:case2xx}, we calculate the variance by substituting $\ti_2=\ti_1$ yielding
\begin{equation}\label{eq:pr:0xxvar}
\var{\interf}=\cov{\interf}{\interf}=\big(\txpd\,\expect{\fadsq{x}{\ti}}-\txpd^2\big)\,\dens\,\int_{\R^2}\plossx^2\,\,\dd x\:.
\end{equation}
Dividing~\eqref{eq:pr:0xxcov} by~\eqref{eq:pr:0xxvar} yields the result.
\end{IEEEproof}

\textbf{Remark:}
\begin{itemize}
\item Since the node locations are not considered in Theorem~\ref{th:case0xx}, mobility is not taken into account in the result.
\item The correlation $\cor{\interf_{\ti_1}}{\interf_{\ti_2}}=0$ for all cases $(1,j,k)$.
\end{itemize}

%% file: src.tex
\section{Analysis of the auto-correlation of interference}\label{sec:src}
Using the general expressions derived in the previous section, we can now analyze the temporal correlation of interference for the three sources of correlation. We start by treating these correlation sources individually and afterwards look at some combinations that provide interesting insights. All plots have been compared to simulation results, which showed a good match. We refrain from plotting the simulation results as they would crowd the figures without providing any additional insights.

\subsection{Correlation by node locations}
The locations of the interfering nodes introduce a correlation that can be intuitively interpreted in the following way: If a receiver has close-by interferers, it is more likely to be disturbed in receiving a message than if no interferers are close-by. If there is no mobility at all, this correlation is independent of the time lag $\delay$, i.e., $\cor{\interf_{\ti_1}}{\interf_{\ti_1+\delay_1}}=\cor{\interf_{\ti_1}}{\interf_{\ti_1+\delay_2}}$ for all positive $\delay_1, \delay_2\in\N$. 

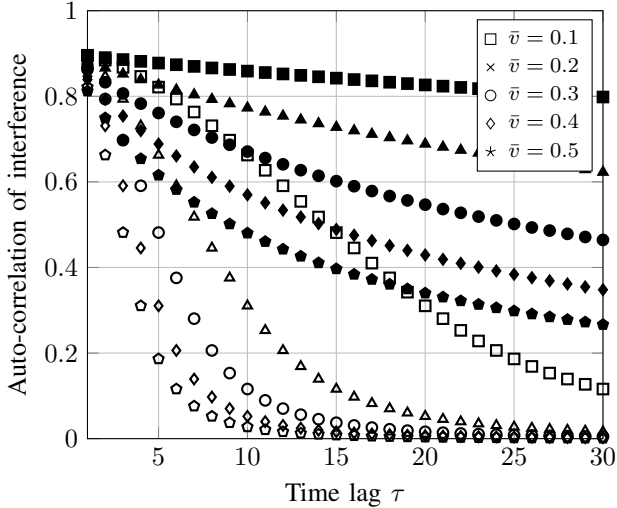
\begin{figure}[tb]
\centering
\begin{tikzpicture}
\begin{axis}[xlabel={Time lag $\delay$},ylabel={Auto-correlation of interference},ymin=0,ymax=1,xmin=1,xmax=30,grid=both,
legend style={at={(0.98,0.98)}, anchor=north east, font=\footnotesize},
legend cell align=left
]

\addlegendimage{mark=square,only marks}
\addlegendimage{mark=x,only marks}
\addlegendimage{mark=o,only marks}
\addlegendimage{mark=diamond,only marks}
\addlegendimage{mark=star,only marks}

\addplot plot[color=black,only marks,mark=square,mark options=solid,style=thick]  table[x index=0,y index=1]  {case201LMM.txt};\addlegendentry{ ~$\speed=0.1$ }
\addplot plot[color=black,only marks,mark=triangle,mark options=solid,style=thick]  table[x index=0,y index=2]  {case201LMM.txt};\addlegendentry{ ~$\speed=0.2$ }
\addplot plot[color=black,only marks ,mark=o,mark options=solid,style=thick]  table[x index=0,y index=3]  {case201LMM.txt};\addlegendentry{ ~$\speed=0.3$ }
\addplot plot[color=black,only marks ,mark=diamond,mark options=solid,style=thick]  table[x index=0,y index=4]  {case201LMM.txt};\addlegendentry{ ~$\speed=0.4$ }
\addplot plot[color=black,only marks  ,mark=pentagon,mark options=solid,style=thick]  table[x index=0,y index=5]  {case201LMM.txt};\addlegendentry{ ~$\speed=0.5$ }

\addplot plot[color=black,only marks ,mark=square*,mark options=solid,style=thick]  table[x index=0,y index=1]  {case201BM.txt};
\addplot plot[color=black,only marks ,mark=triangle*,mark options=solid,style=thick]  table[x index=0,y index=2]  {case201BM.txt};
\addplot plot[color=black,only marks ,mark=*,mark options=solid,style=thick]  table[x index=0,y index=3]  {case201BM.txt};
\addplot plot[color=black,only marks ,mark=diamond*,mark options=solid,style=thick]  table[x index=0,y index=4]  {case201BM.txt};
\addplot plot[color=black,only marks  ,mark=pentagon*,mark options=solid,style=thick]  table[x index=0,y index=5]  {case201BM.txt};

\end{axis}
\end{tikzpicture}
\caption{Auto-correlation function for the case $(2,0,1)$ for the linear mobility model (open marks) and for Brownian motion (filled marks) with different average speeds $\speed$. The speed is measured in meters per time slot and the sending probability is $\txp=0.9$.}
\label{fig:case201}
\end{figure}

In case of mobility, the interference correlation decreases with time~\cite{gong11:icc} depending on the average speed $\speed$ and the type of mobility. 
Fig.~\ref{fig:case201} shows the temporal correlation over the time lag $\delay$ for both linear mobility and Brownian motion. For the same average speed $\speed$, the distance traveled after time $\delay$ is on average smaller in case of Brownian motion, and hence the correlation decreases slower with $\delay$. In general, the decrease of correlation depends only on the distance traveled during the time lag $\delay$ or its distribution in case it is random (e.g. for Brownian motion).

\subsection{Correlation by wireless channel}\label{sec:case021}
The wireless channel is modeled as a block fading channel with length $\chlen$ slots. This means that the channel gain due to multi-path propagation from a potential interferer stays unchanged for $\chlen$ slots and then changes to a stochastically independent value. This change is independent for each of the potential interferers. Hence, in each slot on average the channels of $\frac{1}{\msglen}$ interferers change to a new state.
This assumption introduces a correlation to the interference values of different slots for $\chlen>1$. In the expressions for interference correlation in Theorems~\ref{th:case2xx} and~\ref{th:case0xx}, the effect of the channel of node $x$ is covered by the term $\expect{\fad{x}{\ti_1}\fad{x}{\ti_2}}$.
For Nakagami fading, this term depends on the fading parameter $\nakm$, the channel block length $\chlen$, and the time lag $\delay$. It is given by
\begin{equation}\label{eq:channel}
\expect{\fad{x}{\ti_1}\fad{x}{\ti_2}}=
\begin{cases}
\frac{\nakm+1}{\nakm}-\frac{\delay}{\nakm\chlen} & \mbox{for }\delay<\chlen\\
1 & \mbox{for }\delay\geq\chlen.
\end{cases}
\end{equation}
In the special case of Rayleigh fading ($\nakm=1$), this term simplifies to $\expect{\fad{x}{\ti_1}\fad{x}{\ti_2}}=2-\frac{\delay}{\chlen}$ for $\delay<\chlen$.

\textbf{Remark:} In case a node travels at least half the wavelength during a single time slot, the channel can be assumed stochastically independent for consecutive slots, i.e., $\chlen=1$. Under this assumption fading does not contribute to interference correlation and equations for cases $(i,1,k)$ for some $i,j\in\{0,1,2\}$ apply.

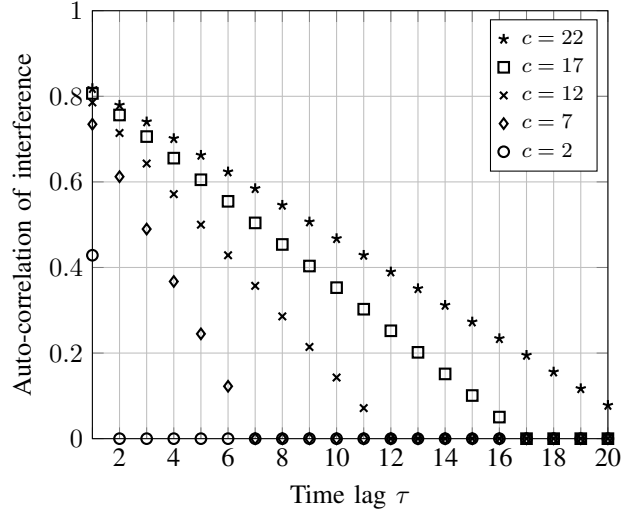
\begin{figure}[tb]
\centering
\begin{tikzpicture}
\begin{axis}[xlabel={Time lag $\delay$},ylabel={Auto-correlation of interference},ymin=0,ymax=1,xmin=1,xmax=20,grid=both,xtick={1,2,3,4,5,6,7,8,9,10,11,12,13,14,15,16,17,18,19,20},
            xticklabels={,2,,4,,6,,8,,10,,12,,14,,16,,18,,20},
legend style={at={(0.98,0.98)}, anchor=north east, font=\footnotesize},
legend cell align=left
]

\addplot plot[color=black,mark=star,only marks,style=thick]  table[x index=0,y index=5]  {case021.txt};\addlegendentry{ ~$\chlen=22$ }
\addplot plot[color=black,mark=square,only marks,style=thick]  table[x index=0,y index=4]  {case021.txt};\addlegendentry{ ~$\chlen=17$ }
\addplot plot[color=black,mark=x,only marks,style=thick]  table[x index=0,y index=3]  {case021.txt};\addlegendentry{ ~$\chlen=12$ }
\addplot plot[color=black,mark=diamond,only marks,style=thick]  table[x index=0,y index=2]  {case021.txt};\addlegendentry{ ~$\chlen=7$ }
\addplot plot[color=black,mark=o,only marks,style=thick]  table[x index=0,y index=1]  {case021.txt};\addlegendentry{ ~$\chlen=2$ }

\end{axis}
\end{tikzpicture}
\caption{Auto-correlation function for the case $(0,2,1)$ for varying the value of the channel block length $\chlen$. The Nakagami fading parameter is assumed to be $\nakm=\frac{1}{2}$ and the sending probability is $\txp=0.9$.}
\label{fig:case021}
\end{figure}

\begin{figure*}
\centering
\begin{tikzpicture}
	\draw [color=black!20,fill=black!5,rounded corners=3pt] (-0.7,2.65) rectangle ++(1.4,0.3);
	\draw [color=black!20,fill=black!5,rounded corners=3pt] (0.9,2.65) rectangle ++(1.4,0.3);
	\draw [fill=black!10,rounded corners=3pt] (2.5,2.65) rectangle ++(1.4,0.3);
	\draw [color=black!20,fill=black!5,,rounded corners=3pt] (4.1,2.65) rectangle ++(1.4,0.3);
	\draw [color=black!20,fill=black!5,rounded corners=3pt] (5.7,2.65) rectangle ++(1.4,0.3);
	\draw [fill=black!10,rounded corners=3pt] (7.3,2.65) rectangle ++(1.4,0.3);
	\draw [color=black!20,fill=black!5,rounded corners=3pt] (8.9,2.65) rectangle ++(1.4,0.3);
	\draw [color=black!20,fill=black!5,rounded corners=3pt] (10.5,2.65) rectangle ++(1.4,0.3);
	\node [align=center, text width=2cm] at (3.2,2.8) {\scriptsize $\ti_1$};
	\node [align=center, text width=2cm] at (8,2.8) {\scriptsize $\ti_2$};

	\draw (-0.8,-0.1) rectangle ++(1.6,0.3);
	\node [align=center, text width=2cm] at (0,0.05) {\scriptsize $1$};
	\draw (0.8,-0.1) rectangle ++(1.6,0.3);
	\node [align=center, text width=2cm] at (1.6,0.05) {\scriptsize $2$};
	\draw (2.4,-0.1) rectangle ++(1.57,0.3);
	\node [align=center, text width=2cm] at (3.2,0.05) {\scriptsize $3$};

	\draw (-0.80,0.35) rectangle ++(1.6,0.3);
	\node [align=center, text width=2cm] at (0,0.5) {\scriptsize $1$};
	\draw (0.8,0.35) rectangle ++(1.6,0.3);
	\node [align=center, text width=2cm] at (1.6,0.5) {\scriptsize $2$};
	\draw (2.4,0.35) rectangle ++(1.6,0.3);
	\node [align=center, text width=2cm] at (3.2,0.5) {\scriptsize $3$};

	\draw (-0.8,0.8) rectangle ++(1.6,0.3);
	\node [align=center, text width=2cm] at (0,0.95) {\scriptsize $1$};
	\draw (0.8,0.8) rectangle ++(1.6,0.3);
	\node [align=center, text width=2cm] at (1.6,0.95) {\scriptsize $2$};
	\draw (2.4,0.8) rectangle ++(1.6,0.3);
	\node [align=center, text width=2cm] at (3.2,0.95) {\scriptsize $3$};

	\draw (7.2,0.8) rectangle ++(1.6,0.3);
	\node [align=center, text width=2cm] at (8,0.95) {\scriptsize $1$};
	\draw (8.8,0.8) rectangle ++(1.6,0.3);
	\node [align=center, text width=2cm] at (9.6,0.95) {\scriptsize $2$};
	\draw (10.4,0.8) rectangle ++(1.6,0.3);
	\node [align=center, text width=2cm] at (11.2,0.95) {\scriptsize $3$};

	\draw (5.6,0.35) rectangle ++(1.6,0.3);
	\node [align=center, text width=2cm] at (6.4,0.5) {\scriptsize $1$};
	\draw (7.2,0.35) rectangle ++(1.6,0.3);
	\node [align=center, text width=2cm] at (8,0.5) {\scriptsize $2$};
	\draw (8.8,0.35) rectangle ++(1.6,0.3);
	\node [align=center, text width=2cm] at (9.6,0.5) {\scriptsize $3$};

	\draw (4.03,-0.1) rectangle ++(1.57,0.3);
	\node [align=center, text width=2cm] at (4.8,0.05) {\scriptsize $1$};
	\draw (5.6,-0.1) rectangle ++(1.6,0.3);
	\node [align=center, text width=2cm] at (6.4,0.05) {\scriptsize $2$};
	\draw (7.2,-0.1) rectangle ++(1.6,0.3);
	\node [align=center, text width=2cm] at (8,0.05) {\scriptsize $3$};

	\draw (0.8,1.25) rectangle ++(1.6,0.3);
	\node [align=center, text width=2cm] at (1.6,1.4) {\scriptsize $1$};
	\draw (2.4,1.25) rectangle ++(1.6,0.3);
	\node [align=center, text width=2cm] at (3.2,1.4) {\scriptsize $2$};
	\draw (4,1.25) rectangle ++(1.57,0.3);
	\node [align=center, text width=2cm] at (4.8,1.4) {\scriptsize $3$};

	\draw (0.8,1.7) rectangle ++(1.6,0.3);
	\node [align=center, text width=2cm] at (1.6,1.85) {\scriptsize $1$};
	\draw (2.4,1.7) rectangle ++(1.6,0.3);
	\node [align=center, text width=2cm] at (3.2,1.85) {\scriptsize $2$};
	\draw (4,1.7) rectangle ++(1.6,0.3);
	\node [align=center, text width=2cm] at (4.8,1.85) {\scriptsize $3$};
	
	\draw (7.2,1.7) rectangle ++(1.6,0.3);
	\node [align=center, text width=2cm] at (8,1.85) {\scriptsize $1$};
	\draw (8.8,1.7) rectangle ++(1.6,0.3);
	\node [align=center, text width=2cm] at (9.6,1.85) {\scriptsize $2$};
	\draw (10.4,1.7) rectangle ++(1.6,0.3);
	\node [align=center, text width=2cm] at (11.2,1.85) {\scriptsize $3$};

	\draw (5.63,1.25) rectangle ++(1.57,0.3);
	\node [align=center, text width=2cm] at (6.4,1.4) {\scriptsize $1$};
	\draw (7.2,1.25) rectangle ++(1.6,0.3);
	\node [align=center, text width=2cm] at (8,1.4) {\scriptsize $2$};
	\draw (8.8,1.25) rectangle ++(1.6,0.3);
	\node [align=center, text width=2cm] at (9.6,1.4) {\scriptsize $3$};

	\draw (2.4,2.15) rectangle ++(1.6,0.3);
	\node [align=center, text width=2cm] at (3.2,2.3) {\scriptsize $1$};
	\draw (4,2.15) rectangle ++(1.6,0.3);
	\node [align=center, text width=2cm] at (4.8,2.3) {\scriptsize $2$};
	\draw (5.6,2.15) rectangle ++(1.57,0.3);
	\node [align=center, text width=2cm] at (6.4,2.3) {\scriptsize $3$};

	\draw (7.23,2.15) rectangle ++(1.57,0.3);
	\node [align=center, text width=2cm] at (8,2.3) {\scriptsize $1$};
	\draw (8.8,2.15) rectangle ++(1.6,0.3);
	\node [align=center, text width=2cm] at (9.6,2.3) {\scriptsize $2$};
	\draw (10.4,2.15) rectangle ++(1.6,0.3);
	\node [align=center, text width=2cm] at (11.2,2.3) {\scriptsize $3$};

	\node [align=center] at (12.9,2.75) {\scriptsize Indices};
	\node [align=center] at (12.9,2.3) {\scriptsize $(1,1)$};
	\node [align=center] at (12.9,1.85) {\scriptsize $(2,1)$};
	\node [align=center] at (12.9,1.4) {\scriptsize $(2,2)$};
	\node [align=center] at (12.9,0.95) {\scriptsize $(3,1)$};
	\node [align=center] at (12.9,0.5) {\scriptsize $(3,2)$};
	\node [align=center] at (12.9,0.05) {\scriptsize $(3,3)$};

	\draw [->] (-1,-0.3) -- (12.2,-0.3);
	\node [align=center, text width=2cm] at (11.7,-0.5) {\scriptsize time $\ti$};
	
\end{tikzpicture}
\caption{An illustration of the potential starting slots of two messages covering $\ti_1$ and $\ti_2$. In the shown setup ($\msglen=3$ and $\delay=3$), there are six possibilities, of which the indices are shown on the right hand side.}
\label{fig:txtwo}
\end{figure*}
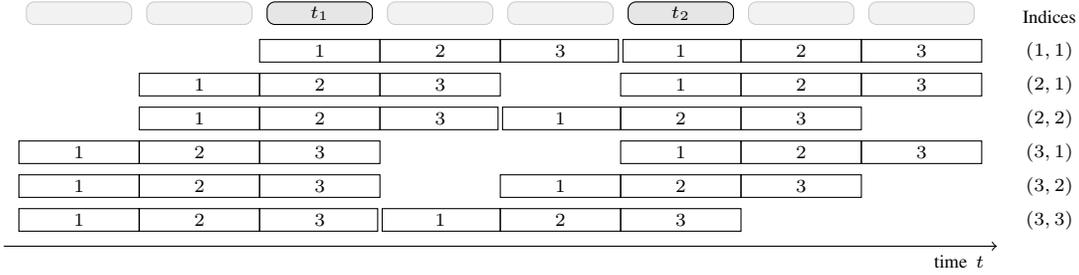

When the wireless channel (i.e., fading) is the only source of interference correlation, we are in a static case $(0,2,1)$. 
A plot of interference correlation in this case is shown in Fig.~\ref{fig:case021} over the time lag $\delay$ for different values of the channel block length $\chlen$. The correlation decreases linearly with the time lag $\delay$ and vanishes for all $\delay\geq\chlen$. For a given time lag, slower fading (higher values of $\chlen$) implies a higher correlation. In the limit for a constant channel, i.e., $\chlen\to\infty$, we get 
\begin{equation}
	\lim_{\chlen\to\infty}\cor{\interf_{\ti}}{\interf_{\ti+\delay}}=\frac{\txp}{1+\nakm-\nakm\txp}
\end{equation}	
independent of $\delay$.

\subsection{Correlation by data traffic}

\begin{figure}[b!]
\centering
\begin{tikzpicture}
	\draw [color=black!20,fill=black!5,rounded corners=3pt] (-0.7,0.85) rectangle ++(1.4,0.3);
	\draw [fill=black!10,rounded corners=3pt] (0.9,0.85) rectangle ++(1.4,0.3);
	\node [align=center, text width=2cm] at (1.6,1) {\scriptsize $\ti_1$};
	\draw [color=black!20,fill=black!5,rounded corners=3pt] (2.5,0.85) rectangle ++(1.4,0.3);
	\draw [fill=black!10,rounded corners=3pt] (4.1,0.85) rectangle ++(1.4,0.3);
	\node [align=center, text width=2cm] at (4.8,1) {\scriptsize $\ti_2$};
	\draw [color=black!20,fill=black!5,rounded corners=3pt] (5.7,0.85) rectangle ++(1.4,0.3);

	\draw (-0.8,-0.1) rectangle ++(1.6,0.3);
	\node [align=center, text width=2cm] at (0,0.05) {\scriptsize $1$};
	\draw (0.8,-0.1) rectangle ++(1.6,0.3);
	\node [align=center, text width=2cm] at (1.6,0.05) {\scriptsize $2$};
	\draw (2.4,-0.1) rectangle ++(1.6,0.3);
	\node [align=center, text width=2cm] at (3.2,0.05) {\scriptsize $3$};
	\draw (4,-0.1) rectangle ++(1.6,0.3);
	\node [align=center, text width=2cm] at (4.8,0.05) {\scriptsize $4$};

	\draw (0.8,0.35) rectangle ++(1.6,0.3);
	\node [align=center, text width=2cm] at (1.6,0.5) {\scriptsize $1$};
	\draw (2.4,0.35) rectangle ++(1.6,0.3);
	\node [align=center, text width=2cm] at (3.2,0.5) {\scriptsize $2$};
	\draw (4,0.35) rectangle ++(1.6,0.3);
	\node [align=center, text width=2cm] at (4.8,0.5) {\scriptsize $3$};
	\draw (5.6,0.35) rectangle ++(1.6,0.3);
	\node [align=center, text width=2cm] at (6.4,0.5) {\scriptsize $4$};

	\draw [->] (-1,-0.3) -- (7.5,-0.3);
	\node [align=center, text width=2cm] at (7,-0.5) {\scriptsize time $\ti$};
	
\end{tikzpicture}
\caption{An illustration of the slots in which transmissions start that each span over both slots $\ti_1$ and $\ti_2$. In the shown setup ($\msglen=4$ and $\delay=2$), there are two possibilities: messages starting at $\ti_1$ having indices $(1,3)$ and messages starting at $\ti_1-1$ with indices $(2,4)$.}
\label{fig:txone}
\end{figure}
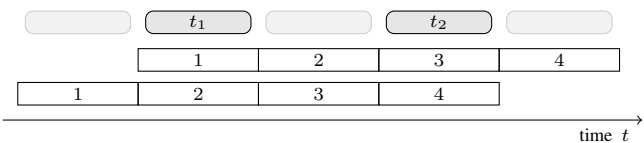

Correlated traffic is caused by having $\msglen>1$, which impacts interference correlation via the expected value $\expect{\tx{x}{\ti_1}\,\tx{x}{\ti_2}}$.

\begin{lemma}[Probability of a node sending in two given time slots]\label{lem:traffic}
The probability $x$ sends in both slots $\ti_1$ and $\ti_2$ is
\begin{eqnarray}
\lefteqn{\expect{\tx{x}{\ti_1}\,\tx{x}{\ti_2}}=\max\big(0,\txp\,(\msglen-\delay)\big)+\frac{\txp^2}{1-\txp(\msglen-1)}}\\\nonumber
&&\sum_{i=0}^{\min(\delay-1,\msglen-1)}  \sum_{j=1}^{\min(\delay-i,\msglen)} \sum_{k=0}^{\left\lfloor\frac{\gap}{\msglen}\right\rfloor}
	\binom{\gap-k\msglen+k}{k}\\\nonumber
&&q^{\gap-k\msglen}(1-q)^k\:,
\end{eqnarray}
where $\gap=\delay-i-j$ and $q=1-\frac{\txp}{1-\txp(\msglen-1)}$.
\end{lemma}
\begin{IEEEproof}
Let us assume that a certain node $x$ is sending in both slots $\ti_1$ and $\ti_2$. Then, there are two possibilities: (${\rm I}$) a single message could span both slots or (${\rm II}$) two different messages are transmitted in these two slots.
A message consists of $\msglen$ time slots; we reference each of them by an index ranging over $1,2,\dots,\msglen$. Let $i$ denote the index of the message at time $\ti_1$ and $j$ the index at time $\ti_2$, and write the indices as tuples $(i,j)$.

The probability $\prob{\txone{\ti_1,\ti_2}}$ that a single message spans over both $\ti_1$ and $\ti_2$ is given by
\begin{equation}\label{eq:pr:ptxone}
\prob{\txone{\ti_1,\ti_2}}=
\begin{cases}
\txp\, (\msglen-\delay) & \mbox{for }\msglen>\delay\\
0 & \mbox{else.}
\end{cases}
\end{equation}
This happens, as shown in Fig.~\ref{fig:txone}, because in each time slot a fraction of $\txp$ nodes start a transmission. For $\msglen>\delay$, there are $\msglen-\delay$ slots for which a message starting there would span both time slots of interest. In this case, the indices $(i,j)$ are always differing by the time lag $j-i=\delay$. For $\msglen\leq\delay$, the time difference between $\ti_1$ and $\ti_2$ is larger than the message length and hence it is impossible that a single message spans both slots.

The probability $\prob{\txtwo{\ti_1,\ti_2}}$ of two different messages being transmitted at slots $\ti_1$ and $\ti_2$ is calculated by summing the probabilities of all possible indices $(i,j)$ (see Fig.~\ref{fig:txtwo}).
The range of the index $i$ is from $\max\big(1,\msglen-(\delay-1)\big)$ to $\msglen$, i.e., if $\delay\geq\msglen$, we have $i=1,\dots,\msglen$; in case $\delay<\msglen$, the index $i$ has to be big enough to avoid a single message spanning both slots $\ti_1$ and $\ti_2$.
The range of the index $j$ depends on the value of $i$, as the message covering slot $\ti_2$ must not overlap with the message covering $\ti_1$. Hence, $j$ ranges from $1$ to $\min\big(\delay-(\msglen-i),\msglen\big)$, i.e., if there is enough space between $\ti_1$ and $\ti_2$, $j$ can go up to $\msglen$; otherwise its maximum value is determined by the case where the two messages are transmitted directly one after another, as for the indices $(1,1)$, $(2,2)$, and $(3.3)$ in Fig.~\ref{fig:txtwo}.

In order to calculate the probability of the situation described by an indices $(i,j)$, we calculate the number of slots between the end of the message covering $\ti_1$ and the beginning of the message covering $\ti_2$ by
\begin{equation}
\gap=\delay-(\msglen-i)-j\:.
\end{equation}
These intermediate slots can be covered by additional messages in case there is enough space, i.e., if $\gap\geq\msglen$. The number $k$ of messages fitting these intermediate slots is at most $k\leq \left\lfloor \frac{\gap}{\msglen}\right\rfloor$ slots, where $\left\lfloor x \right\rfloor$ denotes the biggest integer that is smaller than or equal to $x$. If $k$ messages are present in the intermediate slots, then there are $e=\gap-k\msglen$ slots unoccupied.
The probability of a message starting is, as mentioned in Sec.~\ref{sec:model}, $1-q=\frac{\txp}{1-\txp(\msglen-1)}$, while the probability of an empty slot is $q=1-\frac{\txp}{1-\txp(\msglen-1)}$.
Therefore, the probability that $\ti_1$ and $\ti_2$ are occupied by different messages is
\begin{eqnarray}\label{eq:pr002:s2ss}
\lefteqn{\prob{\txtwo{\ti_1,\ti_2}}=\frac{\txp^2}{1-\txp(\msglen-1)}\sum_{i=\max(1,\msglen-(\delay-1))}^\msglen \hspace{-5mm} \sum_{j=1}^{\min(\delay-(\msglen-i),\msglen)}}
	\\\nonumber
&&\sum_{k=0}^{\left\lfloor\frac{\gap}{\msglen}\right\rfloor}\binom{e+k}{k}\left(\frac{\txp}{1-\txp(\msglen-1)}\right)^k\,
	\left(1-\frac{\txp}{1-\txp(\msglen-1)}\right)^e\:.
\end{eqnarray}

Overall, we can sum the two probabilities calculated above to get the expected value $\expect{\tx{x}{\ti_1}\,\tx{x}{\ti_2}}=\prob{\txone{\ti_1,\ti_2}}+\prob{\txtwo{\ti_1,\ti_2}}$. In~\eqref{eq:pr002:s2ss} we substitute $i$ by $i-\msglen$ to get the result.
\end{IEEEproof}

\begin{corollary}[Simplification for $\delay\leq\msglen$]\label{cor:trafficsimp}
In the case the lag $\delay$ is smallter than or equalt to the message length, the result of Lemma~\ref{lem:traffic} simplifies to
\begin{equation}
\expect{\tx{x}{\ti_1}\,\tx{x}{\ti_2}}=\txp\,(\msglen-\delay)
+\txp\,\frac{\delay(1-\txpc)+\txpc^{\delay+1}-\txpc}{1-\txpc}
\end{equation}
where $\txpc=1-\frac{\txp}{1-\txp(\msglen-1)}$.
\end{corollary}
\begin{IEEEproof}
If we consider the assumption $\delay\leq\msglen$ in~\eqref{eq:pr:ptxone} only the first case can occur. In~\eqref{eq:pr002:s2ss} the upper bounds of the first two sums simplify to $\delay-1$ and $\delay-i$, respectively. In the third sum the upper bound ${\left\lfloor\frac{\gap}{\msglen}\right\rfloor}=0$ and hence there is only one summand with $k=0$. Thus, we have $\binom{e+k}{k}\,\left(\frac{\txp}{1-\txp(\msglen-1)}\right)^k=1$ and overall we have
\begin{eqnarray}
\expect{\tx{x}{\ti_1}\,\tx{x}{\ti_2}}&=&\txp\,(\msglen-\delay)
+\frac{\txp^2}{1-\txp(\msglen-1)}\\\nonumber
&&\sum_{i=0}^{\delay-1}  \sum_{j=1}^{\delay-i}
	\left(1-\frac{\txp}{1-\txp(\msglen-1)}\right)^{\delay-i-j}\:.
\end{eqnarray}
The inner sum of this expression is a geometric series (with the power being $0,\dots,\delay-i-1$). After replacing the closed form result of the inner sum, the outer sum also results in a geometric series but with the first term ($i=0$) missing.
Applying the sum expression of geometric series twice yields
\begin{eqnarray}
\sum_{i=0}^{\delay-1}  \sum_{j=1}^{\delay-i}
	\txpc^{\delay-i-j}&=&\sum_{i=0}^{\delay-1}  
		\frac{1-\txpc^{\delay-i}}{1-\txpc}\\\nonumber
		&=&\frac{\delay-\sum_{i=0}^{\delay-1}\txpc^{\delay-i}}{1-\txpc}\\\nonumber
&=&	\frac{\delay-\frac{1-\txpc^{\delay+1}}{1-\txpc}+1}{1-\txpc}\:,
\end{eqnarray}
where the $1$ is due to the sum starting at $i=1$ instead of $0$. Applying some basic algebra leads to the result.
\end{IEEEproof}

\begin{figure}[tb]
\centering
\begin{tikzpicture}[scale=0.9]
\begin{axis}[
		xlabel=Time lag $\tau$,
		xmin=1,
		xmax=21,
		ymin=1,
		ymax=11,
		xtick={1,2,...,20},
		xticklabels={1,,3,,5,,7,,9,,11,,13,,15,,17,,19},
		mesh/cols=20,
		mesh/rows=40,
		ylabel=Message length $d$,
		colormap={bw}{gray(0cm)=(1); gray(1cm)=(0)},
    colorbar,
		xticklabel style = {xshift=0.18cm},
    yticklabel style = {yshift=0.3cm},
    view={0}{90}]
		
    \addplot3[surf,shader=flat] file {case002heat2.txt};
\end{axis}
\end{tikzpicture}
\caption{Auto-correlation function for the case $(0,0,2)$ for varying the value of the message length $\msglen$. The sending probability is $\txp=0.05$.}
\label{fig:case002cont}
\end{figure}
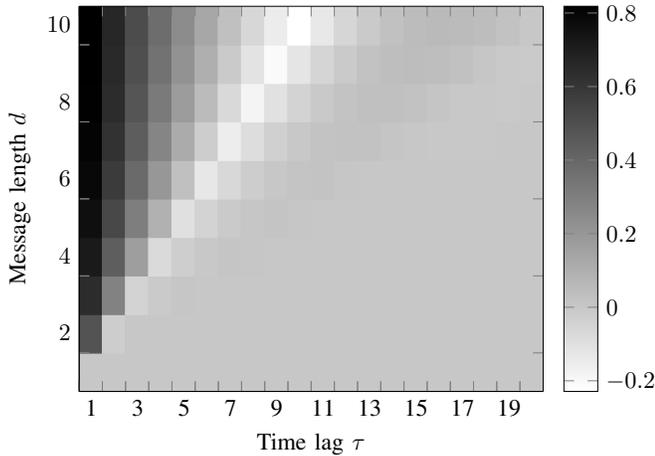

We investigate the temporal correlation of interference when the traffic is the only source of correlation (case $(0,0,2)$) with the aid of 
Fig.~\ref{fig:case002cont}. It shows a heat map of the interference auto-correlation for different message lengths $\msglen$. Correlation is highest for $\delay=1$ and decreases until the lag matches the message length ($\delay=\msglen$), where it is negative. For lags above $\msglen$ it increases to reach a small positive value, from where an oscillating behavior with reducing amplitude is observed. The lag for which zero crossings exist depend, besides the message length, on the sending probability $\txp$: for higher $\txp$ the correlation is in general smaller which implies that it reaches zero for smaller $\delay$ and it gets more negative at $\delay=\msglen$.

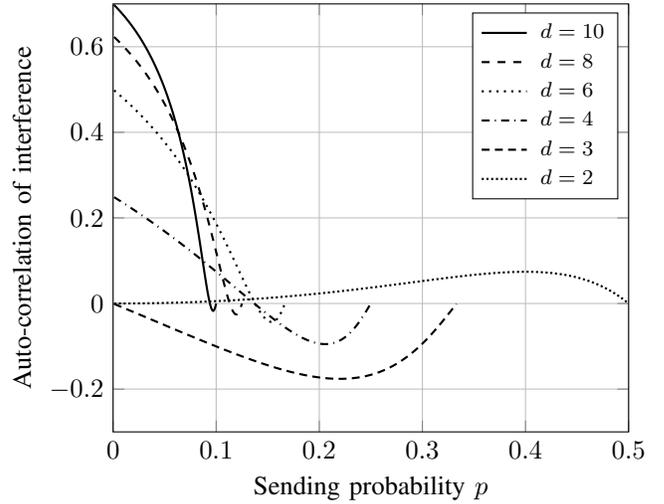
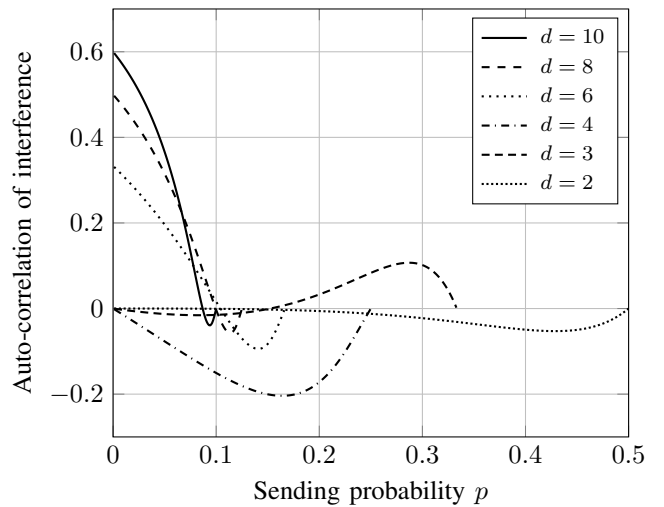
\begin{figure}[!ht]
\centering
  \subfigure[The time lag is $\delay=3$.]{
		\begin{tikzpicture}
			\begin{axis}[xlabel={Sending probability $\txp$},ylabel={Auto-correlation of interference},ymin=-0.3,ymax=0.7,xmin=0,xmax=0.5,grid=both,
									legend style={at={(0.98,0.98)}, anchor=north east, font=\footnotesize},
									legend cell align=left]
					\addplot plot[color=black,solid,no marks,style=thick]  table[x index=0,y index=9]  {case002p.txt};\addlegendentry{ ~$\msglen=10$ }
					\addplot plot[color=black,dashed,no marks,style=thick]  table[x index=0,y index=7]  {case002p.txt};\addlegendentry{ ~$\msglen=8$ }
					\addplot plot[color=black,dotted,no marks,style=thick]  table[x index=0,y index=5]  {case002p.txt};\addlegendentry{ ~$\msglen=6$ }
					\addplot plot[color=black,dashdotted,no marks,style=thick]  table[x index=0,y index=3]  {case002p.txt};\addlegendentry{ ~$\msglen=4$ }
					\addplot plot[color=black,densely dashed,no marks,style=thick]  table[x index=0,y index=2]  {case002p.txt};\addlegendentry{ ~$\msglen=3$ }
					\addplot plot[color=black,densely dotted,no marks,style=thick]  table[x index=0,y index=1]  {case002p.txt};\addlegendentry{ ~$\msglen=2$ }
\end{axis}
\end{tikzpicture}
\label{fig:case002p1}
  }

  \subfigure[The time lag is $\delay=4$.]{
\begin{tikzpicture}
\begin{axis}[xlabel={Sending probability $\txp$},ylabel={Auto-correlation of interference},ymin=-0.3,ymax=0.7,xmin=0,xmax=0.5,grid=both,
										legend style={at={(0.98,0.98)}, anchor=north east, font=\footnotesize},
										legend cell align=left]
					\addplot plot[color=black,solid,no marks,style=thick]  table[x index=0,y index=9]  {case002p2.txt};\addlegendentry{ ~$\msglen=10$ }
					\addplot plot[color=black,dashed,no marks,style=thick]  table[x index=0,y index=7]  {case002p2.txt};\addlegendentry{ ~$\msglen=8$ }
					\addplot plot[color=black,dotted,no marks,style=thick]  table[x index=0,y index=5]  {case002p2.txt};\addlegendentry{ ~$\msglen=6$ }
					\addplot plot[color=black,dashdotted,no marks,style=thick]  table[x index=0,y index=3]  {case002p2.txt};\addlegendentry{ ~$\msglen=4$ }
					\addplot plot[color=black,densely dashed,no marks,style=thick]  table[x index=0,y index=2]  {case002p2.txt};\addlegendentry{ ~$\msglen=3$ }
					\addplot plot[color=black,densely dotted,no marks,style=thick]  table[x index=0,y index=1]  {case002p2.txt};\addlegendentry{ ~$\msglen=2$ }
\end{axis}
\end{tikzpicture}
  \label{fig:case002p2}
 }
	\caption{Auto-correlation function for the case $(0,0,2)$ for varying the sending probability $\txp$ and the value of the message length $\msglen$.}
	\label{fig:case002p}
\end{figure}

A detailed study of the impact of $\txp$ on correlation is presented in the two plots of Fig.~\ref{fig:case002p}.
The first important observation is that the traces can be separated into two groups: The influence of $\txp$ is different for $\msglen\leq\delay$ than for $\msglen>\delay$. For $\msglen\leq\delay$ the correlation is always negative if $\msglen\!\!\mod 2=\delay\!\!\mod 2$ or otherwise mostly positive, only for small $\txp$ it can take small negative values. Furthermore, it converges to zero for $\txp\to 0$.

This behavior can be explained in the following way: Let us assume we have a message length $\msglen=2$. Since on average $\txp$ nodes start a new transmission in each slot and they are chosen from the nodes that are idle, the nodes start to form two groups. One group of nodes start their transmissions in even slots while the other is starting transmissions in odd slots. This group formation is stronger for higher $\txp$. Hence, we have a negative correlation for even values of $\delay$ as mostly the same nodes transmit in $\ti$ and $\ti+\delay$, while we have positive correlation for odd values of $\delay$ as mostly nodes of different groups are transmitting in these two slots.

For $\msglen>\delay$ there are significantly higher correlation values for small $\txp$ and in the limit for $\txp\to 0$ it approaches $\lim_{\txp\to 0}\cor{\interf_{\ti}}{\interf_{\ti+\delay}}\stackrel{\msglen>\delay}{=}\frac{\msglen-\delay}{\msglen}$. For higher $\txp$ the correlation decreases and becomes negative. 
 
For all cases, we have $\lim_{\txp\to\frac{1}{\msglen}}\cor{\interf_{\ti}}{\interf_{\ti+\delay}}=0$, in which case all nodes are always transmitting and hence there is zero variance and covariance.

\subsection{Correlation by multiple sources}

\subsubsection{Channel and traffic}

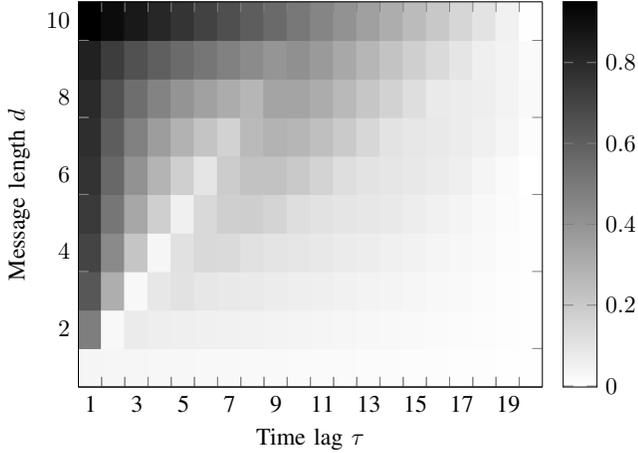
\begin{figure}[tb]
\centering
\begin{tikzpicture}[scale=0.9]
\begin{axis}[
		xlabel=Time lag $\tau$,
		xmin=1,
		xmax=21,
		ymin=1,
		ymax=11,
		xtick={1,2,...,20},
		xticklabels={1,,3,,5,,7,,9,,11,,13,,15,,17,,19},
		mesh/cols=20,
		mesh/rows=40,
		ylabel=Message length $d$,
		colormap={bw}{gray(0cm)=(1); gray(1cm)=(0)},
    colorbar,
		xticklabel style = {xshift=0.18cm},
    yticklabel style = {yshift=0.3cm},
    view={0}{90}]
		
    \addplot3[surf,shader=flat] file {case022heat.txt};
\end{axis}
\end{tikzpicture}
\caption{Auto-correlation function for the case $(0,2,2)$ for varying the value of the message length $\msglen$. The sending probability is $\txp=0.1$, the channel coherence time is $\chlen=22$ and the Nakagami parameter is $m=2$.}
\label{fig:case022cont}
\end{figure}

Fig.~\ref{fig:case022cont} shows a heat map of the auto-correlation when both channel and traffic introduce correlation, which corresponds to case $(0,2,2)$. Results are shown for $\chlen=22$. It can be noted that for given values of $\chlen$ and $\msglen$ the correlation is highest for $\delay=1$ and vanishes for high lags (at least $\delay>\chlen,\msglen$). The correlation vanishes in the limit $\delay\to\infty$. 
For each value of $\msglen$ there is a sharp change of trend at two points: The first is at $\delay=\chlen$ and the second at $\delay=\msglen$. These are the points where the correlation caused by the traffic and by the channel, respectively, are at their minimum. The contribution of the channel is zero at $\delay=\chlen$ and does not change for higher lags, while the contribution of traffic is negative for $\delay=\msglen$ and increases when further increasing $\delay$.

The case $\msglen=10$ is special since all nodes are transmitting all the time (i.e., we have a traffic intensity $\txpd=1$). In such a setting the traffic does not cause any correlation and hence the correlation of interference is fully determined by the channel correlation. This corresponds to the case $(0,2,0)$ for which the correlation shows a linear dependence on $\delay$ (topmost row in the heat map).

\subsubsection{Node locations, channel, and traffic}
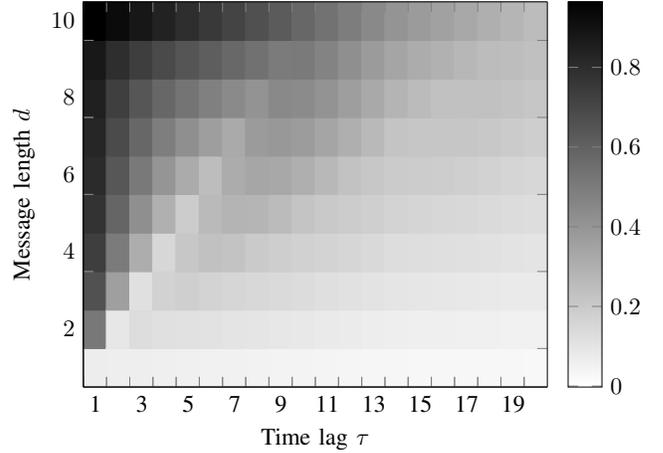
\begin{figure}[tb]
\centering
\begin{tikzpicture}[scale=0.9]
\begin{axis}[
		xlabel=Time lag $\tau$,
		xmin=1,
		xmax=21,
		ymin=1,
		ymax=11,
		xtick={1,2,...,20},
		xticklabels={1,,3,,5,,7,,9,,11,,13,,15,,17,,19},
		mesh/cols=20,
		mesh/rows=40,
		ylabel=Message length $d$,
		colormap={bw}{gray(0cm)=(1); gray(1cm)=(0)},
    colorbar,
		xticklabel style = {xshift=0.18cm},
    yticklabel style = {yshift=0.3cm},
    view={0}{90}]
		
    \addplot3[surf,shader=flat] file {case222heat.txt};
\end{axis}
\end{tikzpicture}
\caption{Auto-correlation function for the case $(2,2,2)$ for varying the value of the message length $\msglen$. The sending probability is $\txp=0.1$ and the block fading length is $\chlen=14$ and $m=1$, i.e., Rayleigh fading is adopted. A static network is considered, i.e., no node mobility, $\speed=0$.}
\label{fig:case222cont}
\end{figure}

When accounting for all three sources of interference correlation and no mobility (case $(2,2,2)$), the auto-correlation evolves as shown in Fig.~\ref{fig:case222cont}. 
Correlation values start at a rather high level for small $\delay$ and decrease for higher $\delay$, although not monotonically, as there are also ranges of $\delay$ where correlation slightly increases. For $\delay$ beyond the message length $\msglen$ and channel coherence time $\chlen$, the correlation approaches its static value as determined by the node locations, i.e., case $(2,0,1)$. In the limit $\delay\to\infty$ it converges to the values of that case. This is explained by noting that the contribution of the node locations to the correlation of interference does not change with $\delay$. Hence, for values of $\delay$ for which the correlation contribution from other sources vanishes, the node locations are the only source of correlation and fully determine its value.

The specific trends of the auto-correlation in the plot are determined by the specific network parameters. The first qualitative change of the trend is the first local minimum, i.e. the point where it first starts to increase, which is located at $\delay=\msglen$. This corresponds to case $(0,0,2)$, where a similar local minimum is present in the plot. The second qualitative change of the trend can be found at $\delay=\chlen=14$, which is the time lag for which the wireless channel's correlation vanishes. Naturally, the order of these two events depends on which of the parameters $\msglen$ and $\chlen$ has a higher value.
Overall, the correlation behaves similar to case $(0,2,2)$, but for high $\delay$ in case $(2,2,2)$ the correlation approaches a constant value while for case $(0,2,2)$ the correlation approaches zero.

%% file: coherence.tex
\section{Coherence Time}\label{sec:coh}
In the same style that the channel coherence time is defined in~\cite{620535}, we define the \textit{interference coherence time} to be the time lag until the auto-correlation function of the interference becomes small and hence the interference becomes stochastically independent from its original value.
\begin{definition}[Interference coherence time]
The interference coherence time $\cohti$ is the minimum time lag $\delay$ such that the auto-correlation is smaller than a threshold $\cohth$, i.e., 
\begin{equation}
\cohti=\min\big\{\delay\in\N\,|\,\cor{\interf_{\ti}}{\interf_{\ti+\delay}}\leq\cohth\big\}\:.
\end{equation}
\end{definition}

\textbf{Remark:} Note that this is a subjective definition since $\cohti$ is a function of $\cohth$, which has to be chosen in accordance to the considered case. The threshold below which the interference can be assumed to be uncorrelated is in general greater than zero. If, for example, a scenario with no mobility and correlation from the node positions is considered (case $(2,j,k)$), correlation will not drop to zero, no matter how high the time lag $\delay$.

In general the coherence time depends on the sources of correlation and the time they need to uncorrelate. In the following we consider them separately to acquire an insight into their individual role. 

\subsection{Impact of traffic on coherence time}
In this section we consider the coherence time $\cohti$ that results from a threshold $\cohth=0$. 
If all transmissions span $\msglen$ slots, the correlation of interference caused by it monotonically decreases for $\msglen$ slots (see Fig.~\ref{fig:case002cont}). However, the coherence time typically is shorter as after $\msglen$ slots the auto-correlation is negative and hence crossed zero earlier, i.e., $\cohti\leq\msglen$. Hence, for the analysis of the coherence time we can adopt the simplified expression from Corollary~\ref{cor:trafficsimp}.

In order to calculate the coherence time, we have to find the value of $\delay$ such that $\cor{\interf_{\ti_1}}{\interf_{\ti_2}}= 0$. For general parameters, there is usually no slot that exactly reaches zero correlation and therefore, we calculate instead the time lag $\delay$ until correlation reaches zero and then round it to the next higher integer. 
\begin{theorem}\label{th:coh002}
The coherence time when traffic is the only source of correlation (case $(0,0,2)$) is
\begin{equation}\label{eq:thcoh002}
\cohti=\left\lceil\frac{\log(1-\txpd)}{\log\txpc}\right\rceil\:,
\end{equation}
where $\txpc=1-\frac{\txp}{1-\txp(\msglen-1)}$ and $\lceil x\rceil=\min\{y\in\N\,|\,y\geq x\}$ is the smallest integer being larger than or equal to $x$.
\end{theorem}
\begin{IEEEproof}
The correlation of interference $\cor{\interf_{\ti_1}}{\interf_{\ti_2}}$ is defined as in Theorem~\ref{th:case0xx}.
Since traffic is the only source of correlation, we substitute $\expect{\fad{x}{\ti_1}\fad{x}{\ti_2}}=\expect{\fadsq{x}{\ti}}=1$. Further, we substitute the result of Corollary~\ref{cor:trafficsimp} for $\expect{\tx{x}{\ti_1}\,\tx{x}{\ti_2}}$ into Theorem~\ref{th:case0xx} yielding
\begin{equation}
\cor{\interf_{\ti_1}}{\interf_{\ti_2}}=\frac{\txp\,(\msglen-\delay)
+\txp\,\frac{\delay(1-\txpc)+\txpc^{\delay+1}-\txpc}{1-\txpc}-\txpd^2}{\txpd-\txpd^2}\:.
\end{equation}
Solving the equation $\cor{\interf_{\ti_1}}{\interf_{\ti_2}}=0$ for the time lag $\delay$ gives
\begin{equation}\label{eq:pr:coh002c}
\delay=\frac{\log(1-\txpd)}{\log\txpc}\:.
\end{equation}
In general the solution of $\delay$ in this expression is a non-integer and hence we have to round it to the next higher integer as the correlation is monotonically decreasing with $\delay$ and we aim for a correlation being smaller or equal to zero.
\end{IEEEproof}

\textbf{Remark:} Although Theorem~\ref{th:coh002} is derived for no fading, the same expression holds for fading with $c=1$, i.e., for case $(0,1,2)$ as the only consequence of considering fading is to divide the auto-correlation function by a constant value $\frac{\nakm+1}{\nakm}$.

\begin{figure}[tb]
\centering
\begin{tikzpicture}
\begin{axis}[xlabel={Sending probability $\txp$},ylabel={Coherence time $\cohti$},ymin=1,ymax=10,xmin=0,xmax=0.5,grid=both,
legend style={at={(0.98,0.98)}, anchor=north east, font=\footnotesize},
legend cell align=left]

\addplot plot[color=black,solid,no marks,style=thick]  table[x index=0,y index=9]  {coh002.txt};\addlegendentry{ ~$\msglen=10$ }
\addplot plot[color=black,dashed,no marks,style=thick]  table[x index=0,y index=7]  {coh002.txt};\addlegendentry{ ~$\msglen=8$ }
\addplot plot[color=black,dotted,no marks,style=thick]  table[x index=0,y index=5]  {coh002.txt};\addlegendentry{ ~$\msglen=6$ }
\addplot plot[color=black,dashdotted,no marks,style=thick]  table[x index=0,y index=3]  {coh002.txt};\addlegendentry{ ~$\msglen=4$ }
\addplot plot[color=black,densely dotted,no marks,style=thick]  table[x index=0,y index=1]  {coh002.txt};\addlegendentry{ ~$\msglen=2$ }

\end{axis}
\end{tikzpicture}
\caption{Interference coherence time if considering solely the traffic as source of temporal correlation (case $(0,0,2)$) over the sending probability $\txp$ for varying the message length $\msglen$. In this plot the rounding to the next higher integer is omitted to avoid high overlap of the curves, .i.e., we plot~\eqref{eq:pr:coh002c} instead of~\eqref{eq:thcoh002}.}
\label{fig:coh002}
\end{figure}
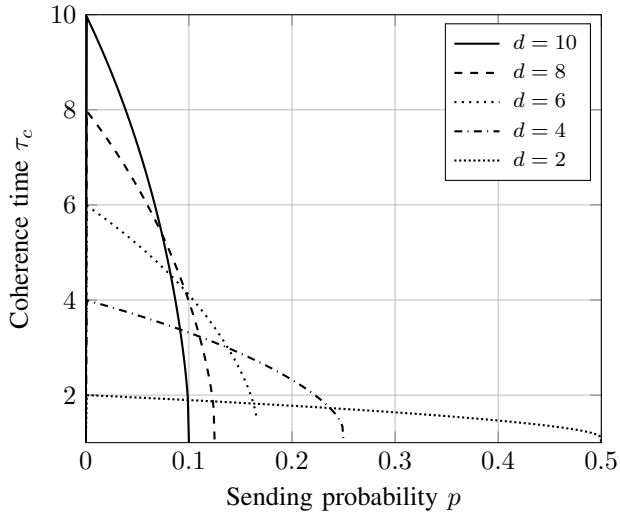

Fig.~\ref{fig:coh002} shows the corresponding plot of the interference coherence time. 
It shows that in case of very small sending probabilities $\txp$ (close to zero) the coherence time is roughly equal to the message length ($\cohti\approx\msglen$). For increasing values of $\txp$ the coherence time is monotonically decreasing and approaches its minimum for $\txpd\to 1$, which is $\lim_{\txp\to\frac{1}{\msglen}}\cohti=1$. Hence, in this case interference is already uncorrelated in consecutive slots.

It is interesting to notice that the coherence time depends on the sending probability $\txp$. In potential applications that require two uncorrelated slots, e.g. a retransmission protocol, the retransmission back-off interval (=time lag) has to be adjusted to the traffic load of the network. For higher traffic loads the back-off interval could be shortened based on this coherence time result, leading to a lower transmission delay at the nodes. There might be of course other reasons that prevent the back-off interval from being too short, but still this example illustrates the potential of having a better understanding of interference dynamics. 

\subsection{Impact of channel on coherence time}
If the channel is the only source of correlation, i.e., case $(0,2,1)$, assuming again $\cohth=0$ we have the following result.
\begin{theorem}
The interference coherence time $\cohti$ if the channel is the only source of correlation (case $(0,2,1)$) equals the channel coherence time $\chlen$. 
\end{theorem}
\begin{IEEEproof}
From Theorem~\ref{th:case0xx} we have that the correlation coefficient for the case $(0,2,1)$ is
\begin{equation}
\cor{\interf_{\ti_1}}{\interf_{\ti_2}}=\frac{\txp\big(\expect{\fad{x}{\ti_1}\fad{x}{\ti_2}}-1\big)}{\frac{(\nakm+1)}{\nakm}-\txp}\:,
\end{equation}
where $\expect{\fad{x}{\ti_1}\fad{x}{\ti_2}}$ is given in~\eqref{eq:channel}.
For the case $\delay=\chlen-1$ we have $\cor{\interf_{\ti_1}}{\interf_{\ti_2}}=\frac{\txp}{\chlen(1+\nakm-\nakm\txp))}$, which is always positive for $\txp>0$. As for $\delay\geq\chlen$ the correlation vanishes, the coherence time is always equal to $\chlen$.
\end{IEEEproof}

\subsection{Impact of node locations on coherence time}
If the node locations are considered as a source of correlation and a threshold $\cohth=0$, it is important to assume mobility. Otherwise the temporal correlation caused by only node locations is constant for all $\delay\geq 1$ and never reaches $\cohth$. In case other sources of correlation exist, the correlation converges to this constant for $\delay\to\infty$ (actually the convergence is rather fast for reasonable values of $\chlen$ and $\msglen$).
Specifically, if the static node locations are the only source of correlation, we have $\cor{\interf_{\ti_1}}{\interf_{\ti_2}}=\txp$~\cite{ganti09:interf-correl,schilcher12:tmc}, independent of $\delay$. In such case it makes no sense to talk about a coherence time as for $\cohth=0$ (or more generally for sufficiently small $\cohth$).

Let us assume that nodes move at an average speed $\speed>0$. The temporal correlation of the interference is monotonically decreasing to its limit $\lim_{\ti\to\infty}\cor{\interf_{\ti_1}}{\interf_{\ti_2}}=0$. For finite time, however, it will get arbitrarily small but remain positive. Hence we choose a threshold $\cohth>0$ for our analysis.

\begin{figure}[tb]
\centering
\begin{tikzpicture}
\begin{axis}[xlabel={Sending probability $\txp$},ylabel={Coherence time $\cohti$},ymin=0,ymax=80,xmin=0,xmax=1,grid=both,
legend style={at={(0.02,0.98)}, anchor=north west, font=\footnotesize},
legend cell align=left]

\addplot plot[color=black,solid,no marks,style=thick]  table[x index=0,y index=1]  {coh201_hr.txt};\addlegendentry{ ~$\speed=0.1$ }
\addplot plot[color=black,dashed,no marks,style=thick]  table[x index=0,y index=2]  {coh201_hr.txt};\addlegendentry{ ~$\speed=0.2$ }
\addplot plot[color=black,dotted,no marks,style=thick]  table[x index=0,y index=3]  {coh201_hr.txt};\addlegendentry{ ~$\speed=0.3$ }
\addplot plot[color=black,dashdotted,no marks,style=thick]  table[x index=0,y index=4]  {coh201_hr.txt};\addlegendentry{ ~$\speed=0.4$ }
\addplot plot[color=black,densely dotted,no marks,style=thick]  table[x index=0,y index=5]  {coh201_hr.txt};\addlegendentry{ ~$\speed=0.5$ }

\end{axis}
\end{tikzpicture}
\caption{Interference coherence time considering the node locations as sole source of temporal correlation (case $(2,0,1)$) with linear mobility. The results are plotted over the sending probability $\txp$ for varying the average speed of the nodes $\speed$ with $\cohth=0.01$. The steps in the plot occur since $\cohti$ is measured in terms of slots and hence is integer.}
\label{fig:coh201}
\end{figure}
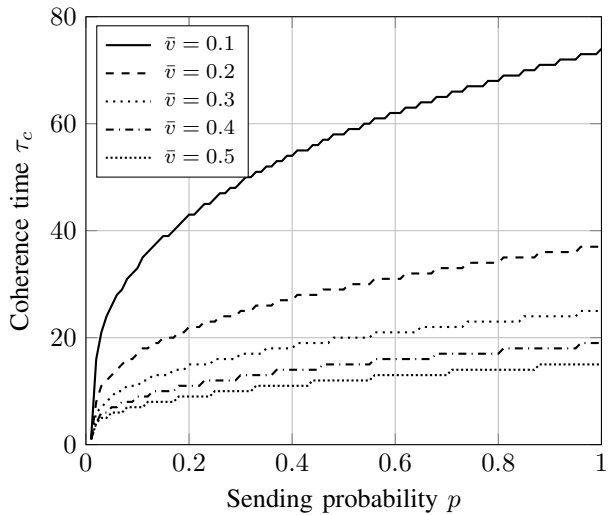

There exists no closed-form expression of the coherence time of interference $\cohti$ when correlation is induced by the node locations.  
This is due to the rightmost integration in~\eqref{eq:cor2xxmob} that to the best of our knowledge has no closed-form solution and therefore, cannot be rearranged to give an expression for $\delay$. Accounting to this, we numerically evaluate the coherence time $\cohti$.
The numerical results are presented in Fig.~\ref{fig:coh201}, where the coherence time $\cohti$ for a threshold $\cohth=0.01$ is investigated. The plot shows $\cohti$ for varying sending probability $\txp$ and different average node speeds $\speed$. The general trend corresponds to intuition: Firstly, $\cohti$ increases with increasing $\txp$, as the temporal correlation of interference increases with $\txp$, making the threshold $\cohth$ to be crossed later. For very small sending probabilities the coherence time is very small (in the limit we have $\lim_{\txp\to 0}\cohti=1$), and the interference in consecutive slots is uncorrelated.
Secondly, a higher average speed $\speed$ of the nodes leads to a smaller coherence time. The reason for it is that with higher speed the nodes reach earlier the distance determining the decorrelation of interference.

The general trends are the same with Brownian motion, but the coherence times are higher, as for this mobility model the distance between the initial and final position of the nodes increases on average slower. This happens because of the back and forth movements of the nodes. We omitted a plot of these results as they do not provide further insights.

%% file: conclusion.tex
\section{Conclusions and future work}\label{sec:con}

We investigated the temporal dynamics of interference in Poisson networks with a focus on its auto-correlation function and coherence time.
We see from the auto-correlation function that the correlation of interference for small time lags typically decreases. This decrease strongly depends on the sources of correlation.
When evaluated over a large span of time lags, the correlation approaches zero, follows a damped oscillation around zero or shows a completely different behavior, all depending on the chosen scenario. It is evident that networking protocols and techniques have to be designed and parametrized differently for such different scenarios in order to operate best. Hence, a good knowledge of the auto-correlation function of interference contributes to improved network performance and robustness.
For example, if interference traces are interpreted as time series, their underlying model order is determined by the auto-correlation function. Hence, our work contributes to model design by providing essential input parameters. It is thus a key enabling result for interference prediction based on time series.

The coherence time of interference can be calculated by means of expressions provided in the article at hand. This is a valuable tool for designing network protocols properly. Moreover, the expressions are simple enough to be implemented on networked nodes, allowing them to adapt to a changing network environment. This is a requirement in scenarios with, e.g., high mobility or high fluctuations in the network nodes.

Ongoing work analyzes how, by estimating a small number of network parameters, the node gathers a good estimate of the interference coherence time. Based on this information, we plan to devise a distributed algorithm to optimize the node's successful communication attempts and thus the overall network performance.

%% file: bios.tex
\begin{IEEEbiography}
    [{\includegraphics[width=1in,height=1.25in,clip,keepaspectratio]{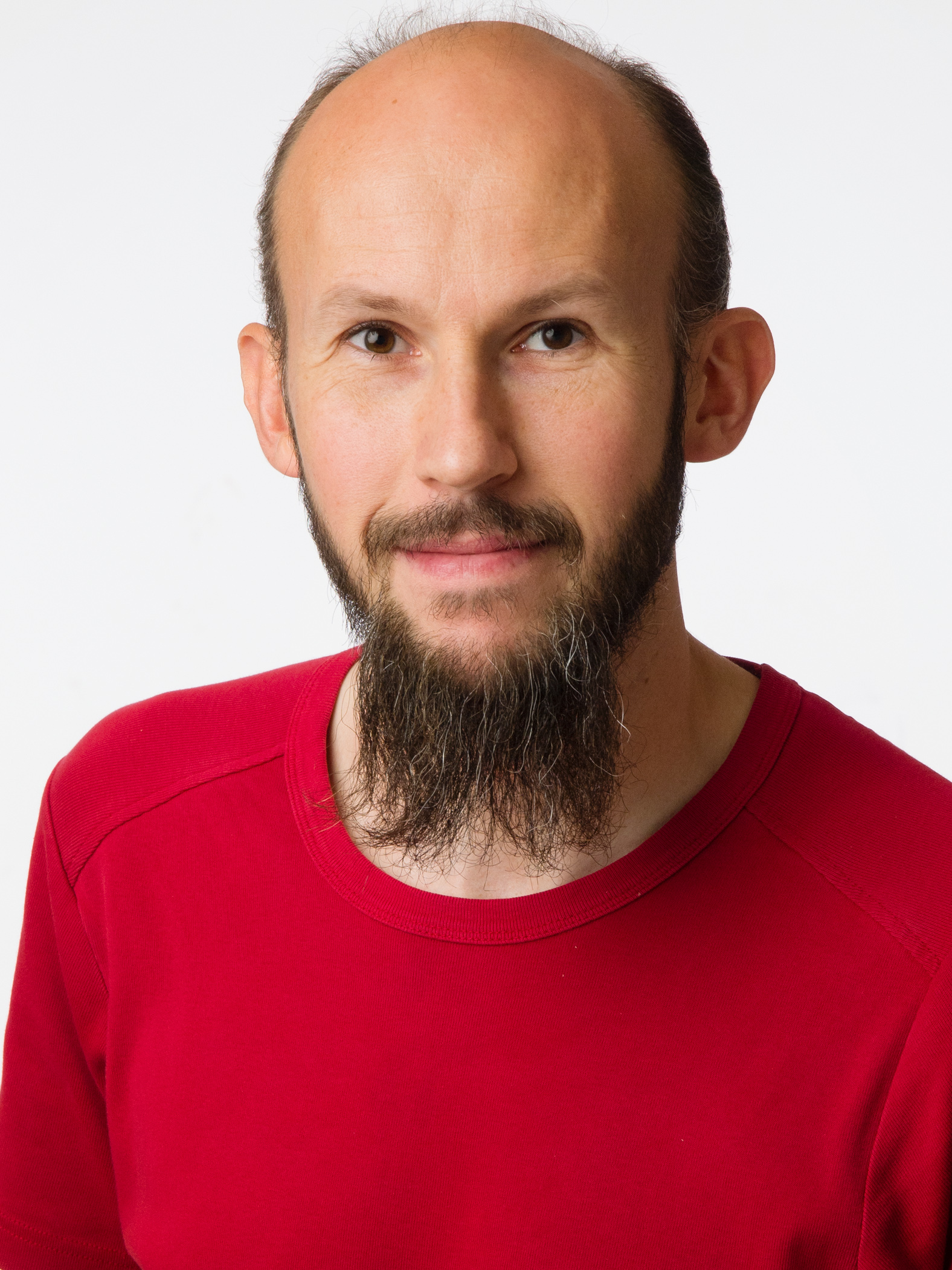}}]{Udo Schilcher}
studied applied computing and
mathematics at the University of Klagenfurt,
where he received two Dipl.-Ing. degrees with
distinction (2005, 2006). Since 2005, he was
research staff member at the Institute of Networked
and Embedded Systems at the University
of Klagenfurt. His doctoral thesis on inhomogeneous
node distributions and interference correlation in
wireless networks and has been awarded with
a Dr. techn. degree with distinction in 2011. 
After his graduation, from 2011 he was Post-Doc,
again at the Institute of Networked
and Embedded Systems at the University
of Klagenfurt.
Since 2016 he is senior researcher
at Lakeside Labs GmbH.
His main interests are interference dynamics
and spatial node distributions in wireless networks,
and stochastic geometry.
He received a best paper
award from the IEEE Vehicular Technology Society.
\end{IEEEbiography}

\begin{IEEEbiography}
    [{\includegraphics[width=1in,height=1.25in,clip,keepaspectratio]{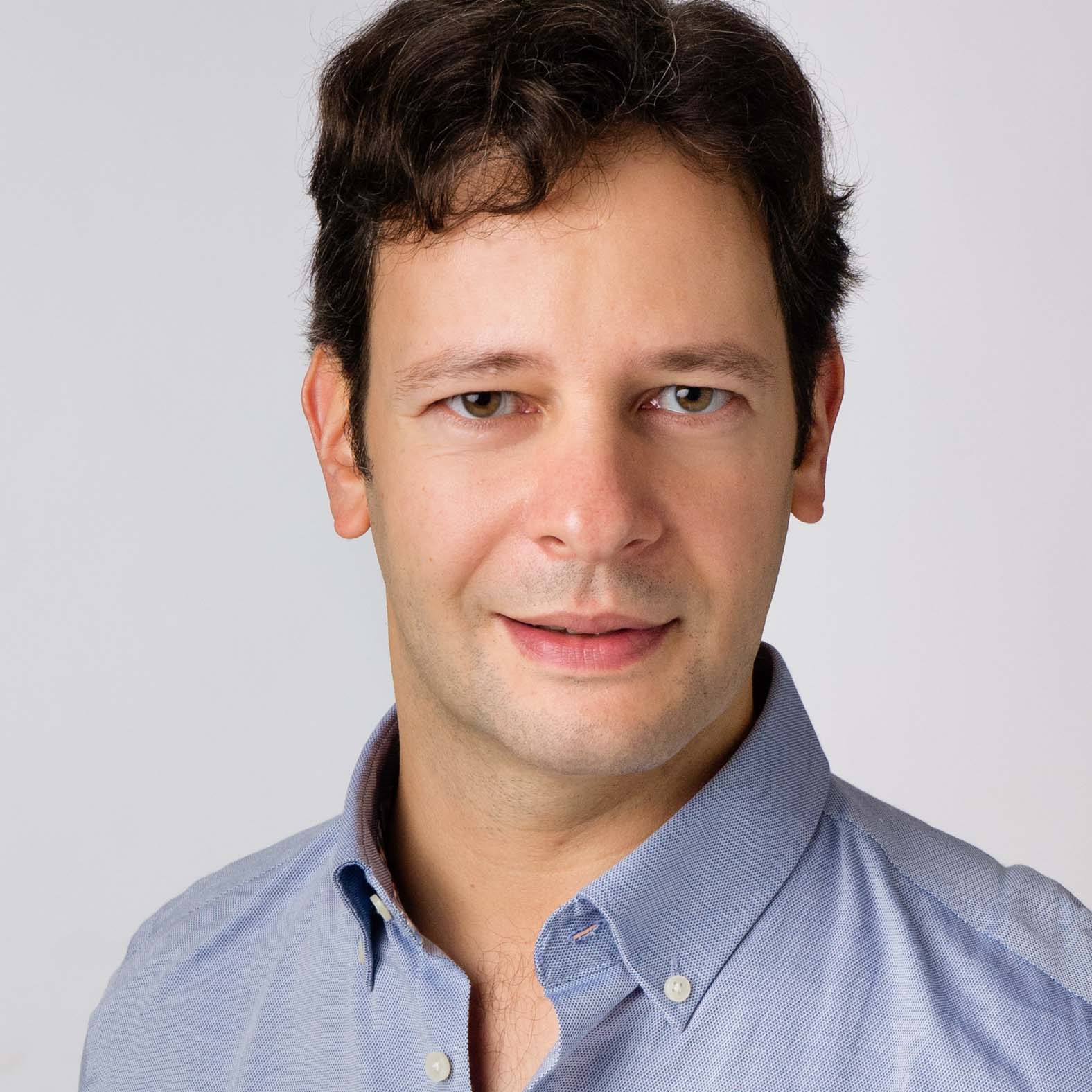}}]{Jorge F. Schmidt}
received the B.Sc. and D.Sc. degrees in electrical engineering from the Universidad Nacional del Sur, Bah\'ia Blanca, Argentina, in 2005 and 2011, respectively. From 2012 to 2014, he was a Postdoctoral Fellow in the Signal Processing and Communications Laboratory at the University of Vigo, Spain. In 2014 he joined the Institute of Networked and Embedded Systems group at University of Klagenfurt, Austria, where he is currently a Senior Researcher. Since 2016 he is also a Senior Researcher at Lakeside Labs GmbH, Austria. His main research interests lie in the area of statistical signal processing and interference modeling and management for wireless communications systems.
He received a best paper award from the ACM SIGSIM.
\end{IEEEbiography}

\begin{IEEEbiography}
    [{\includegraphics[width=1in,height=1.25in,clip,keepaspectratio]{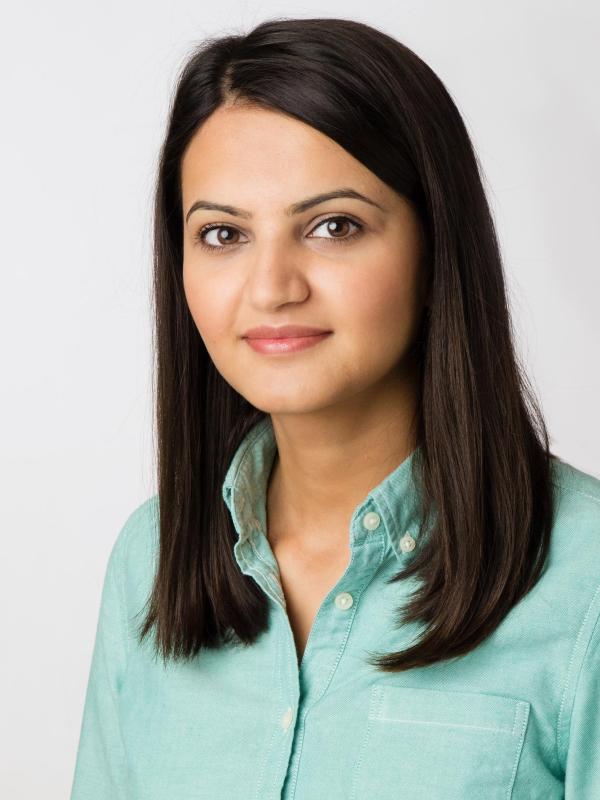}}]{Mahin K. Atiq}
is a research and teaching staff member at the University of Klagenfurt with interests in wireless systems and stochastic modeling. She works toward her doctoral degree in interference prediction in wireless networks. She holds a master degree in information and communication engineering from Sejong University in Seoul (South Korea). 
\end{IEEEbiography}

\begin{IEEEbiography}
    [{\includegraphics[width=1in,height=1.25in,clip,keepaspectratio]{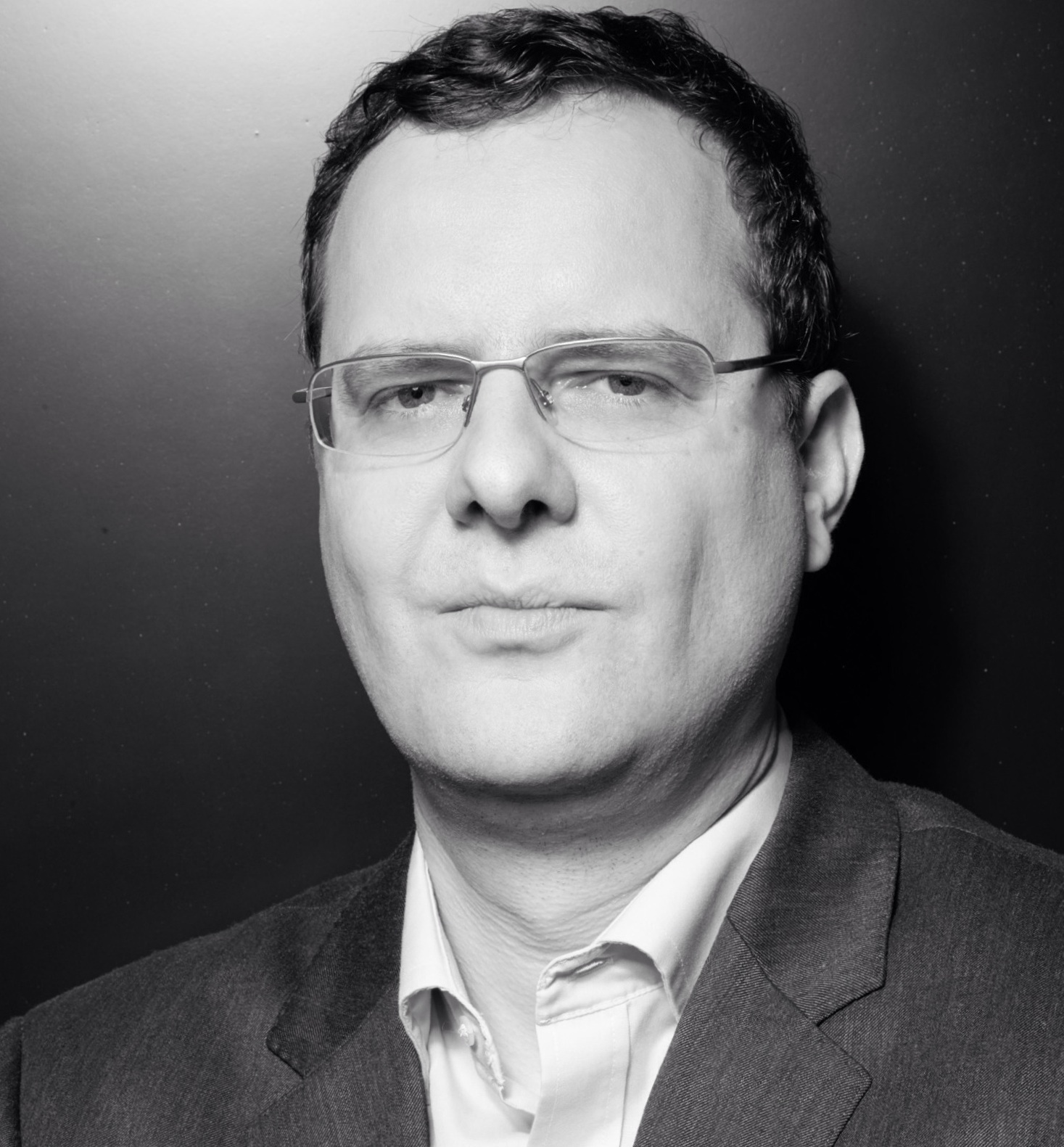}}]{Christian Bettstetter}
(S'98-M'04-SM'09) received the Dipl.-Ing. degree in 1998 and the Dr.-Ing. degree (summa cum laude) in 2004, both in electrical and information engineering from Technische Universit\"at M\"unchen (TUM), Munich, Germany. He was a research and teaching staff member at the Institute of Communication Networks, TUM, until 2003. From 2003 to 2005, he was a senior researcher with DOCOMO Euro-Labs. He has been a professor at the University of Klagenfurt, Austria, since 2005, and founding director of the Institute of Networked and Embedded Systems since 2007. He is also the founding scientific director of Lakeside Labs, a research company on self-organizing networked systems.
\end{IEEEbiography}

%% file: paper.bbl
\begin{thebibliography}{10}

\bibitem{andrews17:mmWave}
J.~G. Andrews, T.~Bai, M.~N. Kulkarni, A.~Alkhateeb, A.~K. Gupta, and R.~W.
  Heath, ``Modeling and analyzing millimeter wave cellular systems,'' {\em
  {IEEE} Trans. Commun.}, vol.~65, pp.~403--430, Jan. 2017.

\bibitem{8115171}
Q.~Cui, X.~Yu, Y.~Wang, and M.~Haenggi, ``The {SIR} meta distribution in
  {Poisson} cellular networks with base station cooperation,'' {\em IEEE
  Transactions on Communications}, vol.~66, pp.~1234--1249, March 2018.

\bibitem{haenggi13:book}
M.~Haenggi, {\em Stochastic Geometry for Wireless Networks}.
\newblock Cambridge University Press, 2013.

\bibitem{schilcher15:tit}
U.~Schilcher, S.~Toumpis, M.~Haenggi, A.~Crismani, G.~Brandner, and
  C.~Bettstetter, ``Interference functionals in {Poisson} networks,'' {\em
  {IEEE} Trans. Inf. Theory}, vol.~62, pp.~370--383, Jan. 2016.

\bibitem{haenggi13:div-poly}
M.~Haenggi and R.~Smarandache, ``Diversity polynomials for the analysis of
  temporal correlations in wireless networks,'' {\em {IEEE} Trans. Wireless
  Commun.}, vol.~12, pp.~5940--5951, Nov. 2013.

\bibitem{tanbourgi14:nakagami}
R.~Tanbourgi, H.~S. Dhillon, J.~G. Andrews, and F.~K. Jondral, ``Dual-branch
  {MRC} receivers under spatial interference correlation and {Nakagami}
  fading,'' {\em {IEEE} Trans. Commun.}, vol.~62, pp.~1830--1844, June 2014.

\bibitem{tanbourgi14:mrc}
R.~Tanbourgi, H.~S. Dhillon, J.~G. Andrews, and F.~K. Jondral, ``Effect of
  spatial interference correlation on the performance of maximum ratio
  combining,'' {\em {IEEE} Trans. Wireless Commun.}, vol.~13, pp.~3307--3316,
  June 2014.

\bibitem{crismani14:tvt}
A.~Crismani, S.~Toumpis, U.~Schilcher, G.~Brandner, and C.~Bettstetter,
  ``Cooperative relaying under spatially and temporally correlated
  interference,'' {\em {IEEE} Trans. Veh. Technol.}, vol.~64, pp.~4655--4669,
  Oct. 2015.

\bibitem{schilcher13:mswim}
U.~Schilcher, S.~Toumpis, A.~Crismani, G.~Brandner, and C.~Bettstetter, ``How
  does interference dynamics influence packet delivery in cooperative
  relaying?,'' in {\em Proc. ACM/IEEE Intern. Conf. on Modeling, Analysis and
  Simulation of Wireless and Mobile Systems (MSWiM)}, (Barcelona, Spain), Nov.
  2013.

\bibitem{7880697}
G.~George, R.~K. Mungara, A.~Lozano, and M.~Haenggi, ``Ergodic spectral
  efficiency in {MIMO} cellular networks,'' {\em IEEE Transactions on Wireless
  Communications}, vol.~16, pp.~2835--2849, May 2017.

\bibitem{atiq17:mswim}
M.~K. Atiq, U.~Schilcher, J.~F. Schmidt, and C.~Bettstetter, ``Semi-blind
  interferenece prediction in wireless networks,'' in {\em Proc. ACM Intern.
  Conf. on Modeling, Analysis and Simulation of Wirel. and Mobile Syst.
  (MSWiM)}, (Miami Beach, FL, USA), pp.~19--23, Nov. 2017.

\bibitem{ganti09:interf-correl}
R.~Ganti and M.~Haenggi, ``Spatial and temporal correlation of the interference
  in {ALOHA} ad hoc networks,'' {\em {IEEE} Commun. Lett.}, vol.~13,
  pp.~631--633, Sept. 2009.

\bibitem{schilcher12:tmc}
U.~Schilcher, C.~Bettstetter, and G.~Brandner, ``Temporal correlation of
  interference in wireless networks with {Rayleigh} block fading,'' {\em {IEEE}
  Trans.~Mobile Comput.}, vol.~11, pp.~2109--2120, Dec. 2012.

\bibitem{6697936}
Y.~Zhong, W.~Zhang, and M.~Haenggi, ``Managing interference correlation through
  random medium access,'' {\em {IEEE} Trans. Wireless Commun.}, vol.~13,
  pp.~928--941, Feb. 2014.

\bibitem{6331038}
M.~Haenggi, ``Diversity loss due to interference correlation,'' {\em {IEEE}
  Commun. Lett.}, vol.~16, pp.~1600--1603, Oct. 2012.

\bibitem{schilcher13:scc}
U.~Schilcher, C.~Bettstetter, and G.~Brandner, ``Temporal correlation of
  interference: Cases with correlated traffic,'' in {\em Proc. ITG Conf. Syst.,
  Commun. and Coding (SCC)}, (Munich, Germany), Jan. 2013.

\bibitem{net:Haenggi09jsac}
M.~Haenggi, J.~G. Andrews, F.~Baccelli, O.~Dousse, and M.~Franceschetti,
  ``Stochastic geometry and random graphs for the analysis and design of
  wireless networks,'' {\em {IEEE} J. Sel. Areas Commun.}, vol.~27,
  pp.~1029--1046, Sept. 2009.

\bibitem{gong14:tmc}
Z.~Gong and M.~Haenggi, ``Interference and outage in mobile random networks:
  Expectation, distribution, and correlation,'' {\em {IEEE} Trans. Mobile
  Comput.}, vol.~13, pp.~337--349, Feb. 2014.

\bibitem{gong11:icc}
Z.~Gong and M.~Haenggi, ``Temporal correlation of the interference in mobile
  random networks,'' in {\em Proc. IEEE Intern. Conf. on Commun. (ICC)},
  (Kyoto, Japan), June 2011.

\bibitem{bettstetter03:tmc}
C.~Bettstetter, G.~Resta, and P.~Santi, ``The node distribution of the random
  waypoint mobility model for wireless ad hoc networks,'' {\em {IEEE} Trans.
  Mobile Comput.}, vol.~2, pp.~257--269, July--September 2003.

\bibitem{620535}
B.~Sklar, ``Rayleigh fading channels in mobile digital communication systems.
  {I}. characterization,'' {\em {IEEE} Commun. Mag.}, vol.~35, pp.~136--146,
  Sept. 1997.

\end{thebibliography}
